\documentclass[fleqn,usenatbib]{mnras}

\usepackage{newtxtext,newtxmath}

\usepackage[T1]{fontenc}
\usepackage{ae,aecompl}

\usepackage{graphicx}	
\usepackage{amsmath}	
\usepackage{amssymb}	
\usepackage{mathtools}

\newcommand{\ddroit}[2]{\ensuremath{\dfrac{\mathrm{d}{#1}}{\mathrm{d}{#2}}}}

\newcommand{\msol}{\ensuremath{\mathrm{M}_\odot}}

\newcommand{\email}[1]{\href{mailto:#1}{\nolinkurl{#1}}}

\title[Tidal disruption of stars in a supermassive black hole binary system]{Tidal disruption of stars in a supermassive black hole binary system: the influence of orbital properties on fallback and accretion rates}

\author[Q. Vigneron et al.]{
Quentin Vigneron,$^{1}$\thanks{E-mail: \email{quentin.vigneron@ens-lyon.fr}}
Giuseppe Lodato,$^{2}$
Alessio Guidarelli$^{2}$
\\
$^{1}$Univ Lyon, ENS de Lyon, CNRS, Centre de Recherche Astrophysique de Lyon UMR5574, F-69230, Saint-Genis-Laval, France\\
$^{2}$Dipartimento di Fisica, Universit\`{a} degli Studi di Milano, Via Celoria 16, 20133 Milano, Italy
}

\date{Accepted XXX. Received YYY; in original form ZZZ}

\pubyear{2016}

\begin{document}
\bibliographystyle{mnras.bst}
\label{firstpage}
\pagerange{\pageref{firstpage}--\pageref{lastpage}}
\maketitle

\begin{abstract}
The disruption of a star by a supermassive black hole generates a sudden bright flare. Previous studies have focused on the disruption by single black holes, for which the fallback rate decays as~$\propto t^{-5/3}$. In this paper, we generalise the study to the case of a supermassive black hole binary (SMBHB), using both analytical estimates and hydrodynamical simulations, looking for specific observable signatures. The range of binary separation for which it is possible to distinguish between the disruption created by a single or a binary black hole concerns typically separations of order a few milliparsecs for a primary of mass $\sim 10^6\msol$. When the fallback rate is affected by the secondary, it undergoes two types interruptions, depending on the initial inclination $\theta$ of the orbit of the star relative to the plane of the SMBHB. For $\theta \lesssim 70^\circ$, periodic sharp interruptions occur and the time of first interruption depends on the distance of the secondary black hole with the debris. If $\theta \gtrsim 70^\circ$, a first smooth interruption occurs, but not always followed by a further recovery of the fallback rate. This implies that most of the TDEs around a SMBHB will undergo periodic sharp interruptions of their lightcurve.
\end{abstract}




\section{Introduction}

When a star wanders too close to a supermassive black hole, it can be disrupted by the strong tidal field created by this compact object. The resulting stellar debris are sent on highly elliptical orbits, or, if their energy is large enough, can even by ejected out of the system. The bound debris accrete onto the black hole, which leads to a flare for which the time dependency of the light curve is generally described by the power law $L(t) \propto t^{-5/3}$ derived by \citet{Rees1988}, \citet{Phinney1989} and \citet{EvansKochanek1989}. Such an event is called a Tidal Disruption Event (TDE).

After many years of surveys since the first detection of such an event by the ROSAT All-Sky survey, a TDE is thought to happen every $10^5$ years in each galaxy \citep{VelzenFarrar2014}. The resulting flare has a very broad spectrum ranging from $\gamma$-rays to X-rays and even down to radio wavebands for the so-called `jetted' TDE, as Swift J1644 \citep{Bloom2011, Burrows2011,Levan2011,Zauderer2011} or in the UV and optical wavebands as in PS1-10jh \citep{Gezari2012}. Simultaneous X-rays and optical flares have also been detected, such as in the case of ASASSN14-li \citep{Holoien2016}. This emission can last from months to years.

Whether the observed lightcurve at some given wavelength should follow or not the time evolution of the fallback rate (and thus produce the signature $t^{-5/3}$ decline) has been the subject of several theoretical investigations \citep{Lodato2011,Shiokawa2015,Guillochon2016,Bonnerot2016}. In addition, even the fact that the fallback rate should simply scale as $t^{-5/3}$ has been questioned, and it has been shown that it may be modified in the early phases, depending on the stellar structure \citep{Lodato2009} or at late times if the disruption is not complete \citep{Guillochon2013}. Here, we will concentrate on the fallback rate and its possible deviations from the power law, expected in the simplest case. So far, these studies generally focused on a TDE around a single black hole. \citet{Liu2009}, \citet{Ricarte2015} and more recently \citet{Coughlin2016} studied the change in the light curve from the power law for a TDE occurring around a supermassive black hole binary (SMBHB). By using ballistic simulations, the first two papers showed that the interaction with the secondary black hole could lead to interruptions in the flare. Flares which present such characteristics could provide a powerful probe of the population of SMBHBs in the Universe. In this regard, \cite{Liu2014} recently proposed a possible candidate of a TDE around a SMBHB.

\citet{Coughlin2016} coupled hydrodynamical and statistical three body simulations to show that the lightcurve could even be totally different from the standard description of TDEs. SMBHBs are expected to form after the coalescence of two galaxies, each having a SMBH in its center. Detecting the presence of a SMBHB at sub-parsec scales is challenging, and most diagnostics (such as a Doppler shift of the broad line region) are ambiguous. Still, these scales are essential to probe from a theoretical perspective, because it is at parsec scales that the process of merging of the binary components is expected to stall \citep{Milosavljevic2001}. Further orbital decay, beyond parsec scales, should take the binary down to milliparsec scales, where the decay time due to gravitational wave emission becomes shorter than the age of the Universe and the binary can then merge and provide a strong source of gravitational waves, that can be detected by upcoming missions, such as LISA.

The purpose of the present paper is to complete both analytically and numerically the work of \citet{Liu2009} and \citet{Ricarte2015}, especially by characterizing with hydrodynamical simulations the light curve of a TDE around a SMBHB exploring a larger parameter space than in the past. Our work also complements that of \citet{Coughlin2016}, in that while they analyze the fallback rate mostly in a statistical sense, we describe more systematically the effects of varying the fundamental physical and geometrical parameters of the system.

The paper is organised as follows. In section $2$ we describe the basic features of a TDE by a single black hole. In Section $3$ we discuss analytically the expectation for the fallback rate around a SMBH binary. In section $4$ we describe our numerical setup. In Section $5$ we show our results. In Section $6$, we discuss the accuracy of the fallback rate computed numerically. In Section $7$ we draw our conclusions.

\section{TDE on a single black hole}
\label{sec::TDE_Single}

In the following, a Newtonian potential is assumed for the black holes. General relativistic effects are essential to determine the fate of the debris as they circularize and possibly form a disc. For the purpose of this paper, in which we are concerned only with the fallback rate, we can safely neglect these effects. 

In order for a star to be disrupted by a black hole, it has to reach a certain distance $R_{\rm t}$ from the black hole, called the tidal radius, and given by:
\begin{eqnarray}
	R_{\rm t} \simeq R_\star\left(\dfrac{M_{\rm h}}{M_\star}\right)^{1/3},
	\label{eq::Rt}
\end{eqnarray}
where $M_{\rm h}$ is the mass of the black hole, $R_\star$ the radius of the star and $M_\star$ the mass of the star. By requiring that $R_{\rm t} > R_{\rm S}$, the Schwarzschild radius of the black hole, the Equation~(\ref{eq::Rt}) leads to a maximum limit of the black hole mass above which no TDE is possible anymore. This limit is $M_{\rm h, max}=5\times10^7 \ \msol$, for non-spinning black holes.

The theory related to a TDE on a single black hole has been developed in the article of \citet{Rees1988} and demonstrated numerically by \citet{EvansKochanek1989}. The accretion rate resulting from the event is associated to the fallback rate of the debris to the pericenter of the initial orbit of the star. This fallback rate is determined by the distribution of orbital energy of the debris by the following relation
\begin{eqnarray}
	\ddroit{M}{T} = \ddroit{M}{E}\dfrac{\left(2\pi GM_{\rm h}\right)^{2/3}}{3} T^{-5/3},
	\label{eq::fallback_rate}
\end{eqnarray}
where $T$ is the orbital period of the bound debris. One of the main assumption behind this formula is that, after the return to pericenter, the debris lose rapidly energy and angular momentum and circularize or accretes to the black hole in a time shorter than $T$. Therefore, we can consider the mass fallback rate to be the accretion rate $\dot{M}$ onto the black hole.

By taking a uniform mass distribution $\text{d}{M}/\text{d}{E}$ we have 
\begin{eqnarray}
	\ddroit{M}{T} = \dfrac{1}{3}\dfrac{M_\star}{t_{\rm min}}\left(\dfrac{t}{t_{\rm min}}\right)^{-5/3},
		\label{eq::dMdT_uni}
\end{eqnarray}
where $t_\text{min}$ corresponds to the return time of the most bound debris and is given by
\begin{eqnarray}
	t_{\rm min} = \dfrac{2\pi R_{\rm t}^3}{(GM_{\rm h})^{1/2}(2R_\star)^{3/2}}.
	\label{eq::tmin}
\end{eqnarray}
This hypothesis of an uniform mass distribution is very simple and far from realistic but has the advantage of giving analytical prediction of the fallback rate which fits well the bolometric luminosity of observed TDEs \citep{Lodato2012}. A better estimate of the energy distribution can be obtained either analytically, by assuming that the stellar structure is unperturbed upon reaching the tidal radius \citep{Lodato2009} or, more realistically, through numerical simulations \citep{Lodato2009,Guillochon2013}. For complete disruptions, even with a more realistic treatment of the energy distribution, the $t^{-5/3}$ decline is preserved at late times, while it may show strong deviations for partial disruptions and/or non parabolic encounters \citep{Guillochon2013}.

\section{Tidal disruption events by a SMBH binary}

We now consider a TDE occurring around a supermassive black hole binary (SMBHB). Such a binary can result from the merging of two galaxies. The population of SMBHBs is thought to be smaller than the population of single supermassive black holes. However, the rate of TDEs detected around a SMBHB is not necessarily lower than for single black holes. For one galaxy, this rate can indeed be at least one order of magnitude higher than the one of a single supermassive black hole \citep[see][]{Amaro2013}.

\subsection{Restriction of the parameters}
\label{sec::choice_parameters}

\begin{figure*}
	\centering
	\includegraphics[width=2\columnwidth]{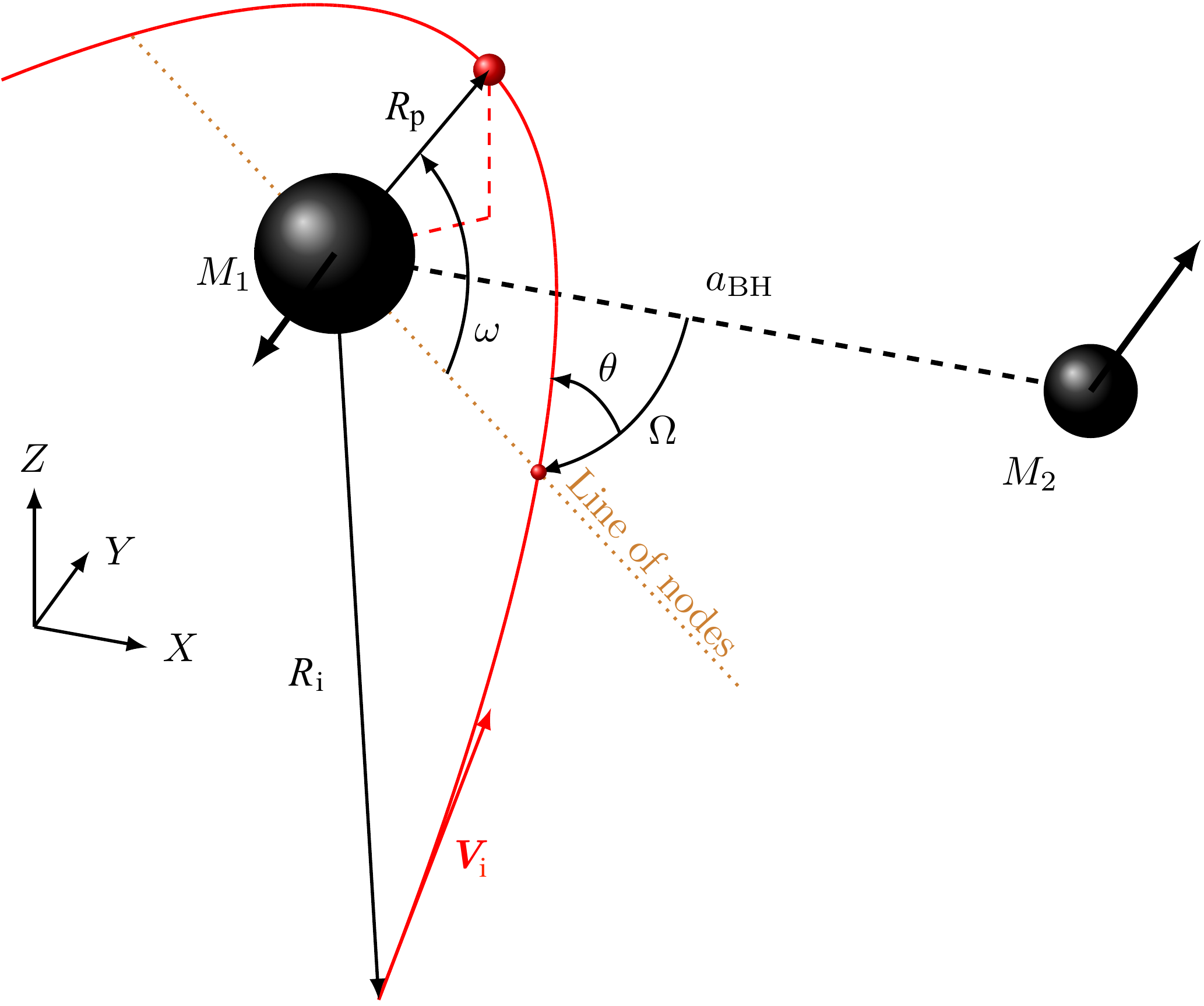}
	\caption{Illustration of the system. The SMBHB is in the $x-y$ plane and has an angular momentum in the direction of positive $z$. The red full line is the initial orbit of the star around the primary. $R_{\rm i}$ and $V_{\rm i}$ are the initial position and velocity of the star.}
	\label{fig::Positions}
\end{figure*}

A TDE around a SMBHB implies many more parameters than on a single black hole. There are two kinds of parameters, those concerning the binary black holes, and those concerning the initial orbit of the star. They are summarized in Figure~\ref{fig::Positions}.

\subsubsection{Binary parameters}

For the binary, we consider the black hole involved in the disruption to be the primary and call $M_1$ its mass. The secondary black hole has a mass $M_2$. We can set the eccentricity $e_{\rm BH}$, the sum of the semi-major axis $a_{\rm BH}$ of the orbit of each black hole around the center of mass, the mass of the primary $M_1$ and the mass ratio $q=M_2/M_1$. \citet{Milosavljevic2001} showed that the eccentricity of SMBHBs is moderated due to a fast circularization after their formation, and with $a_{\rm BH}\lesssim 10$ pc, we can consider the eccentricity to be approximately $e_{\rm BH}=0$. So from now, we consider only SMBHBs with $e_{\rm BH}=0$. Also, in this case the binary separation is constant and equal to $a_{\rm BH}$. We choose to take only moderate black hole masses, with $M_1=10^6\msol$.

The choice of $M_1$, $q$ and $a_{\rm BH}$ is restricted by the theoretical estimations of the fallback rate made later in subsection~\ref{sec::Theo_estimates}.

\subsubsection{Star parameters}
\label{sec::star_parameters}

We take a solar type star of mass $M_\odot$ and radius $R_\odot$. The parameters describing the initial orbit of the star around the primary are the orbital elements: the eccentricity $e_\star$, the pericenter $R_{\rm p}$, the inclination of the orbital plane of the star and that of the binary $\theta$, the angular position of the line of nodes $\Omega$, the apsidal position of the pericenter $\omega$ and the initial true anomaly $\nu_{\rm i}$. Their definition is such that when $\theta=\Omega=\omega=\nu_{\rm i}=0$, the star is at pericenter between the two black holes. The reference plane is the plane of the SMBHB. The reference axis for $\Omega$ is the $x$-axis, i.e. initial axis linking the two black holes.

The true anomaly $\nu_{\rm i}$ is just the initial angular position of the star on the orbit. We can replace, without any loss of generality, $\nu_{\rm i}$ by the initial distance to the primary $R_{\rm i}$. This parameter does not play an essential role in the final result. On the one hand, the star needs to have time to be deformed by the black hole before reaching pericenter, but on the other hand the trajectory between the initial position and the pericenter must not be perturbed by the secondary in order to reach the wanted pericenter. For all the simulations, we took $R_{\rm i}=3 R_{\rm p}$, which satisfies both conditions.

The pericenter is always taken to be at the tidal radius, $R_{\rm p}=R_{\rm t}$. We take in all the simulations a parabolic orbit, i.e. $e_\star=1$. Also, for a single black hole, a study of the effects of an initial elliptic orbit is made by \citet{Bonnerot2015}.

The most interesting parameters are $\theta$, $\Omega$ and $\omega$ because they determine in which direction with respect to the secondary the debris will be thrown. We simulate essentially two different cases which probe two different effects the secondary might have on the debris.

Firstly we make the disruption in the binary plane ($\theta=\omega=0^\circ$) with a range of $\Omega \in \{0^\circ;90^\circ;180^\circ;270^\circ\}$. Because the debris are mainly in the plane, crossing of the secondary into the stream of these debris can occur. Thus we probe the effects of a direct, or at least close, encounter of this black hole with the debris.

Secondly we make the disruption perpendicular the plane of the SMBHB, i.e. $\theta=90^\circ$ and $\omega=90^\circ$, with $\Omega \in \{0^\circ;90^\circ;180^\circ;270^\circ\}$. In this case the secondary never crosses the stream of the debris. These simulations probe the effects of the global modification of the gravitational potential due to the binary. 

Finally we considered a number of intermediate inclinations ($\theta \in \{0^\circ, 10^\circ, 20^\circ, 30^\circ, 40^\circ, 50^\circ, 60^\circ, 70^\circ, 80^\circ, 90^\circ\}$, with $\omega=\Omega=90^\circ$) to understand the transition between the two extreme behaviors.

$\Omega$ is used for probing the influence of the azimuthal position of the stream of the debris with respect to the secondary.

\subsection{Theoretical estimates}
\label{sec::Theo_estimates}

In order to make some estimates of the perturbations the secondary might have on the TDE, one has to consider a three-body system, the two black holes and a debris element. As we will show, the dynamic of the debris depends mainly on the distance between the two black holes.

\subsubsection{Truncation time}
\label{sec::Ttr}

If the binary separation is large enough, the most bound debris are on S-type orbit, i.e. their orbit is not perturbed by the secondary. The fallback rate of such debris is also the classical power law $\dot{M} \propto t^{-5/3}$. Since these debris represent the very first moments of the fallback rate, then the lightcurve will follow the power law in the beginning. However, when arriving to debris that have orbits that can be highly perturbed by the secondary black hole, the power law stops working. In the N-body simulations of \citet{Liu2009} and \citet{Ricarte2015}, the following fallback rate undergoes interruptions. It is however difficult to make analytical predictions for the subsequent orbit of the perturbed debris and for the time at which they will return to pericenter. But it is possible to estimate which debris will be perturbed and which ones will not.

Let us define a critical semi-major axis $a_{\rm cr}$ for the debris orbit above which they will be perturbed by the secondary. Then we can say that the light curve will differ from the classical power law after the time of the first return to pericenter of these perturbed debris. In \citet{Liu2009} they call this time the \textit{truncation time} because they observe net truncations in the light curve. We call it $t_{\rm tr}$. Then using Kepler's third law, we obtain:
\begin{eqnarray}
	t_{\rm tr} = 2\pi\left(\dfrac{a_{\rm cr}^3}{GM_{\rm h}}\right)^{1/2}.
\end{eqnarray}
The critical semi-major axis can have different definitions. The one of \citet{Liu2009} results from a semi-empirical boundary condition between the chaotic behavior or not of a triple system. In this paper we choose \citep[as in][]{Coughlin2016} to take a more usual definition of $a_{\rm cr}$: the size of the Roche lobe $R_{\rm lobe}$. The solution is not analytical and is approximately given by the formula of \citet{Eggleton1983} with a precision of $1\%$:
\begin{eqnarray}
	R_{\rm lobe} = \dfrac{0.49q^{2/3}}{0.6q^{2/3} + \ln{\left(1+q^{1/3}\right)}}a_{\rm BH},
\end{eqnarray}
where $R_{\rm lobe}$ is normalized by the binary separation.

Then $R_{\rm lobe}$ is the maximum apocenter with respect to the primary that a debris can reach without being perturbed. In terms of semi-major axis this gives after some simple algebra:
\begin{eqnarray}
	a_{\rm cr} = \dfrac{R_{\rm lobe} + R_{\rm p}}{2}\approx \dfrac{R_{\rm lobe} }{2},
	\label{eq::acr_L1}
\end{eqnarray}
where the last approximation holds since the typical separation of the binary is of the order of $(M_1/M_\star)^{1/3}$ larger than the tidal radius, see Eq. \ref{eq::a_minmax} below. 

\begin{figure}
	\centering
	\includegraphics[width=\columnwidth]{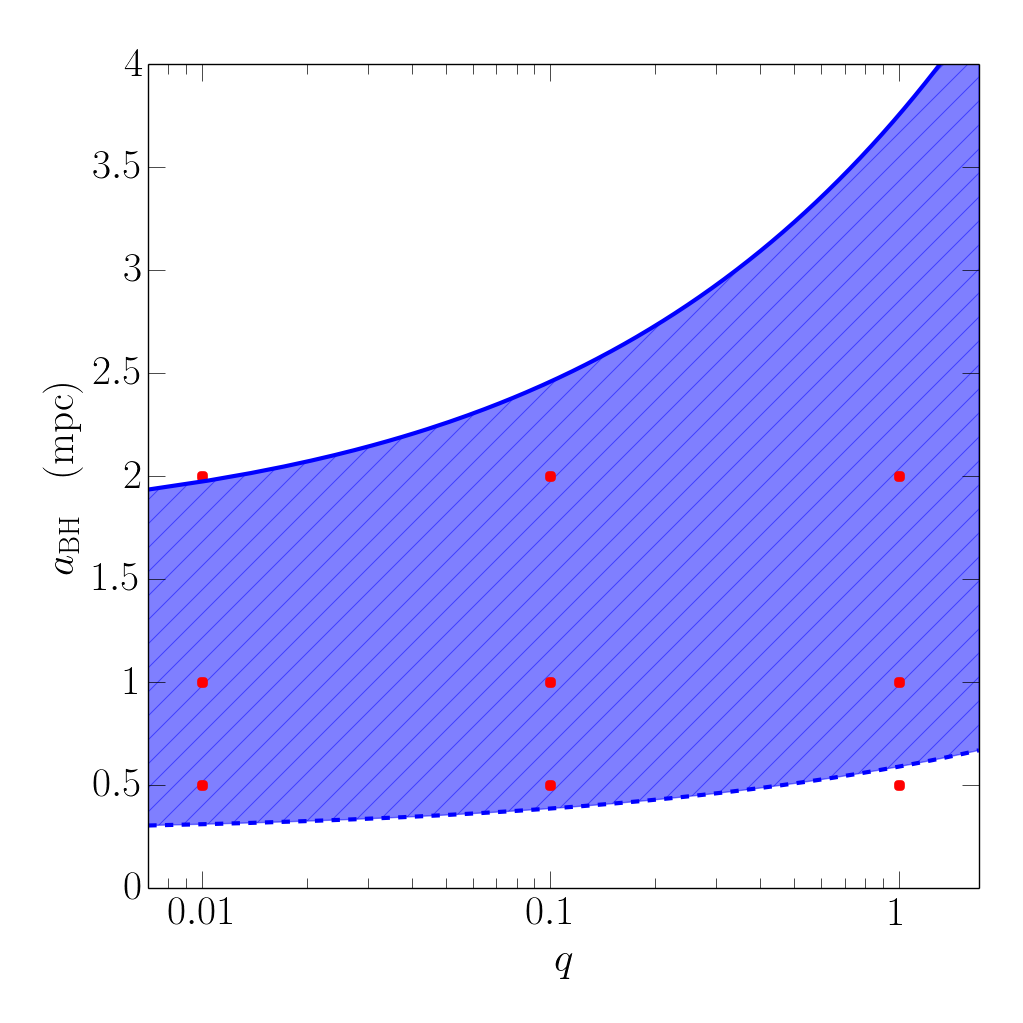}
	\caption{The blue area depicts, as a function of the mass ratio $q$, the range of binary separations for which truncations of the power law can be seen, calculated with Equation~\eqref{eq::a_minmax}. The mass of the primary is $M_1=10^6\msol$. The red dots correspond to the points in the parameter space where the simulations have been done. For each of these points we perform $8$~simulations with different orientations of the initial orbit of the star, $4$~in the plane of the binary and $4$~perpendicular to the plane (see Section~\ref{sec::star_parameters}).}
	\label{fig::a_minmax}
\end{figure}

\subsubsection{Interval of binary separation}
\label{sec::a_minmax}

Here, we estimate the interval of binary separations for which we expect the binary to perturb the TDE lightcurve. As quoted in section~\ref{sec::TDE_Single}, the first return of the debris occurs at $t=t_{\rm min}$. This means that if the truncation time is lower than $t_{\rm min}$, then the power law is not followed at all. Thus we have a lower limit $t_{\rm tr,min}$ for $t_{\rm tr}$. In the same way, we can define an upper limit $t_{\rm tr,max}$ using observational constraints. \citet{Komossa2015} showed that it is possible to follow the evolution of the lightcurve until a luminosity of $\approx 1\%$ of the peak luminosity. This leads to a maximum time of observation of the order of some years. If $t_{\rm tr}$ is above this time, we will not be able to detect any change in the power law and identify the TDE to be around a SMBHB.

These limits of the truncation time can be translated into limits for the binary separation $a_{\rm BH, min}$ and $a_{\rm BH, max}$. This leads to
\begin{eqnarray}
	\begin{dcases}
		a_{\rm BH,min}^{\rm para} = \dfrac{0.6q^{2/3}+\ln{\left(1+q^{1/3}\right)}}{0.49q^{2/3}} R_\star \left(\dfrac{M_1}{M_\star}\right)^{2/3}, \\
		a_{\rm BH,max}^{\rm para} = \epsilon^{-2/5} a_{\rm BH,min}
	\end{dcases}
	\label{eq::a_minmax}
\end{eqnarray}
where $\epsilon$ is the minimum observable luminosity relative to the peak luminosity. We take $\epsilon=0.01$ and $R_{\rm p}=R_{\rm t}$. The superscript ``para'' refers to the fact that in this case we have assumed a parabolic orbit for the star. Roughly, Eq. \ref{eq::a_minmax} implies that $a_{\rm BH, min}$ is the separation at which the binding energy of the binary $\approx GM_1/a_{\rm BH}$ (in the limit that $q\ll1$) is comparable to the energy spread imparted by the tidal disruption onto the stellar debris $\Delta E=GM_1R_\star/R_{\rm t}^2$. Now, it is interesting to note that \citet{Coughlin2016} have shown that in the case of TDEs from a binary black hole, the orbital energy of the incoming star can assume a wide range between $\approx -2GM_1/a_{\rm BH}$ and $\approx 2GM_1/a_{\rm BH}$. It may thus be possible that non-parabolic encounters might alter the range of relevant binary separations significantly. We have thus computed the interval of binary separations for which we expect the fallback to be significantly affected by the binary also for the case in which the incoming star has got a non negligible orbital energy, with modulus $E_\star>0$. We then find:
\begin{eqnarray}
	\begin{dcases}
		a_{\rm BH,min}^{E_\star} = a_{\rm BH,min}^{\rm para}\frac{1}{1\pm E_\star/\Delta E}, \\
		a_{\rm BH,max}^{E_\star} = \epsilon^{-2/5} a_{\rm BH,min}^{E_\star},
	\end{dcases}
	\label{eq::a_minmax2}
\end{eqnarray}
where
\begin{equation}
\Delta E=\frac{GM_1R_\star}{R_{\rm t}^2}\approx \frac{GM_1}{R_\star}\left(\frac{M_1}{M_\star}\right)^{-2/3},
\end{equation}
and the superscript ``$E_\star$'' indicates that here we consider stellar orbits with non-negligible energy.  In Eq. \ref{eq::a_minmax2} the plus and minus sign refer to the case where the stellar orbit is elliptical and hyperbolic, respectively. From \citet{Coughlin2016}, we expect that the maximum $E_\star\approx 2\Delta E$. This would imply a change of at most a factor of a few in the relevant separations for the case of elliptical stellar orbits. For hyperbolic orbits the fallback can be affected by the binary presence already at significantly larger separations. For extremely hyperbolic encounters, when $E_\star>\Delta E$, all the debris become unbound and there will be no fallback flare. In the following, we will consider only the case where $E_\star=0$.

We make use of Equations~(\ref{eq::a_minmax}) to choose the range of binary separations taken in the simulations. Figure~\ref{fig::a_minmax} illustrates the position in the parameter space $(q;a_{\rm BH})$ of the points (in red)  that are simulated, with respect to the interval of binary separation allowed by Equation~(\ref{eq::a_minmax}) (blue area). We choose to take $q<1$, which corresponds to a disruption only on the more massive black hole. \citet{Chen2008,Chen2009} and more recently \citet{Coughlin2016} showed that statistically this is the more probable case. For each point we perform $8$~simulations with different orientations of the initial orbit of the star, 4~in the plane of the binary and 4~perpendicular to the plane (see section~\ref{sec::star_parameters}).

\section{Numerical setup}

We use Smoothed Particle Hydrodynamic (SPH) which is a Lagrangian method of hydrodynamic simulation using fluid particles. In order to take shocks into account we introduce the typical artificial viscosity parameters of SPH, $\alpha^{\rm AV}$ and $\beta^{\rm AV}$, and the switch of \citet{CullenDehnen2010} that reduces $\alpha^{\rm AV}$ and $\beta^{\rm AV}$ away from the shocks. Using this switch, $\alpha^{\rm AV}$ is bounded between a minimum value $\alpha_{\rm min}^{\rm AV}=0$ and a maximum value $\alpha_{\rm max}^{\rm AV}=1$, while $\beta^{\rm AV}=2$.

We perform the simulations using PHANTOM \citep{PriceFederrath2010,LodatoPrice2010,Priceetal2017}. The code units are set to be the solar radius for the distances, the solar mass for the masses and we set $G$ to be equal to $1$. The SMBHB is modeled as a time dependent external force which acts on the particles. In all the simulations we include the gas self-gravity and we use an adiabatic equation of state, with $\gamma = 5/3$. The simulation is initialized by distributing the particles to reproduce the density profile of a polytropic star. This is firstly perform without the external force until the star has reached equilibrium. Then we ensure this star on a parabolic orbit around the primary black hole according to the initial conditions given by the Section~\ref{sec::star_parameters}. Initially the distance $R_i$ of the star to the primary is set to be $3$ times the pericenter distance. Higher values have been tested ($R_{\rm i}=5R_{\rm p}$ for instance) and no discrepancies have been found. One should note that the time for the star to reach pericenter is negligible with respect to the period of the SMBHB. Thereby the trajectory of the star between the initial position and the pericenter is not perturbed by the dynamic of the SMBHB and is the wanted parabola.

In all the simulations we take $N=10^5$ particles. Simulations have been performed also with $N=10^6$ particles and only minor differences have been found (see Section \ref{sec::limits})

In order to avoid the time step to go to zero when particles get very close to one of the black holes, we define an "accretion radius" $R_{\rm acc}$ for both black holes. If a particle passes under this radius it is removed from the simulation. The accretion radius of the secondary is the radius of the Innermost Stable Circular Orbit (ISCO) of this black hole, which is the radius of the last stable orbit, equal to three times the Schwarzschild radius for a non-spinning black hole, such as those considered here.

However, we take another definition for the accretion radius of the primary. For this black hole we take $R_{\rm acc,1}=0.8R_{\rm p}$.  The reason why we do not take the ISCO radius is related to a numerical reason. The part of the stream of debris that is returning to the pericenter is greatly stretched due to the fact that we chose a parabolic orbit. Then very few particles resolve this part of the stream, even with $N=10^6$. It follows that the artificial viscosity is abnormally increased for the particles closest to the black hole, and therefore their dynamics is not well simulated. For instance, with $R_{\rm acc,1}=R_{\rm ISCO,1}$, $R_{\rm ISCO,1}$ being the ISCO radius of the primary, we observed a very fast (and obviously not physical) ejection of some particles after their return to the pericenter. To get rid of this problem, we take a large accretion radius. This is a well known issue when simulating parabolic disruptions \citep{Bonnerot2015} but here we are interested in fallback rate rather than accretion rate, so avoiding this issue by taking a larger accretion radius does not affect our results. 
One should note that this problem is not present for the secondary since we do not have a "parabolic returning", leading to a small density of particles, near this black hole. That is why we take the ISCO radius for its accretion radius.


\section{Results}

\subsection{Influence of the initial orientation}

In this section we analyse the effects of the initial orientation of the orbit of the star at the points of Figure~\ref{fig::a_minmax}, in the case where the initial orbit is in the plane of the binary and in the case where it is perpendicular to this plane.

\subsubsection{Disruption in the plane}
\label{sec::dis_plane}

\begin{figure*}
	\centering
	\includegraphics[width=0.66\columnwidth]{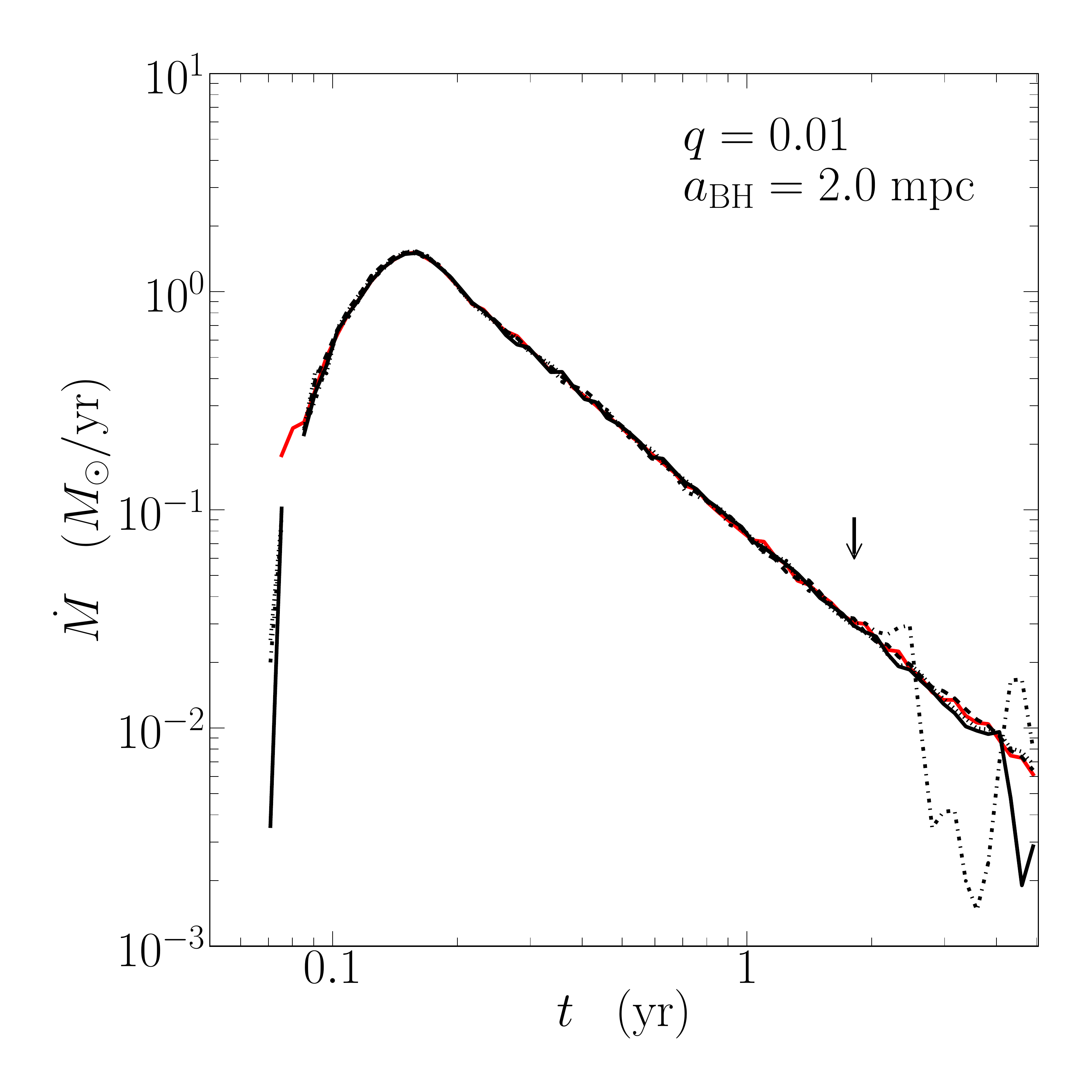}
	\includegraphics[width=0.66\columnwidth]{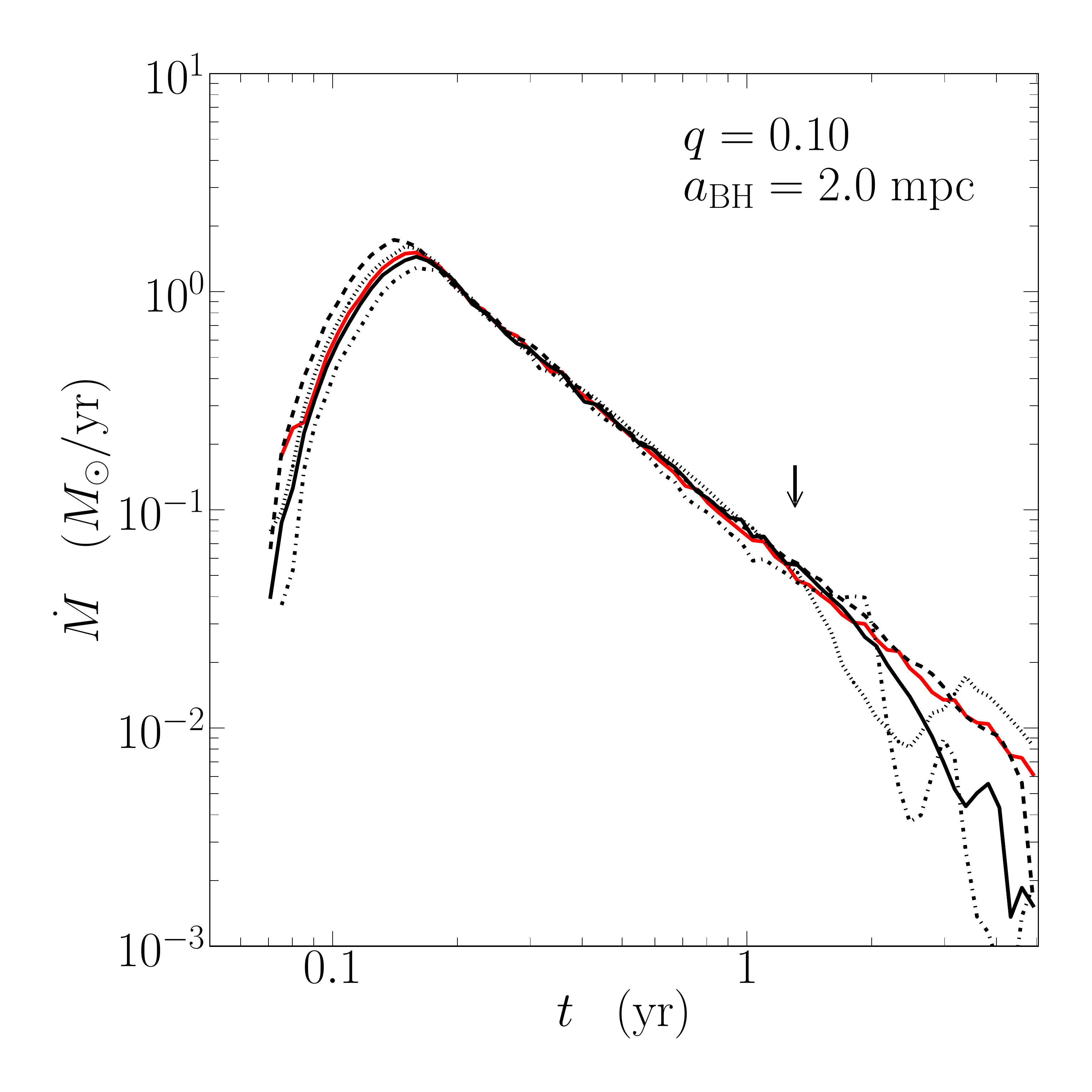}
	\includegraphics[width=0.66\columnwidth]{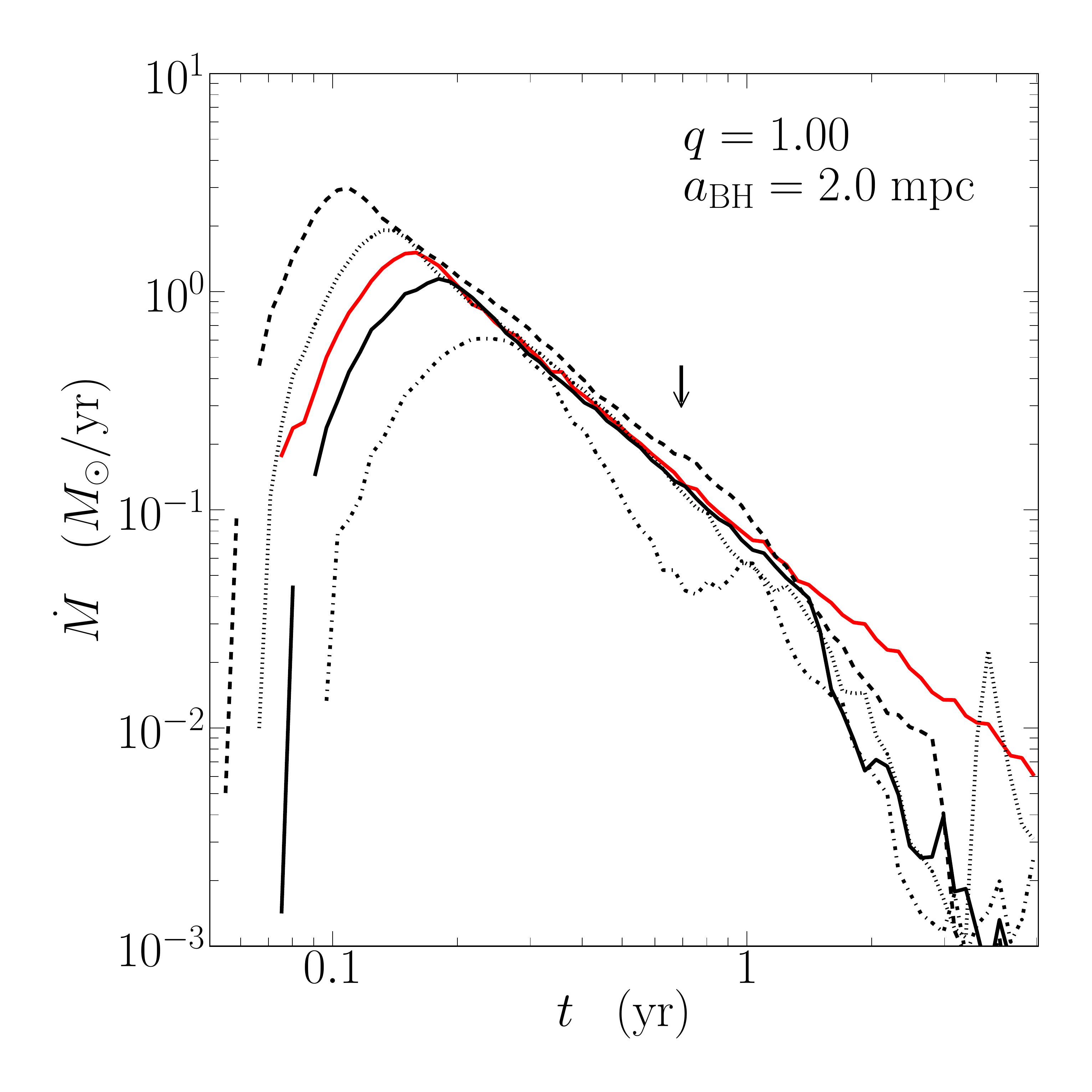}
	\includegraphics[width=0.66\columnwidth]{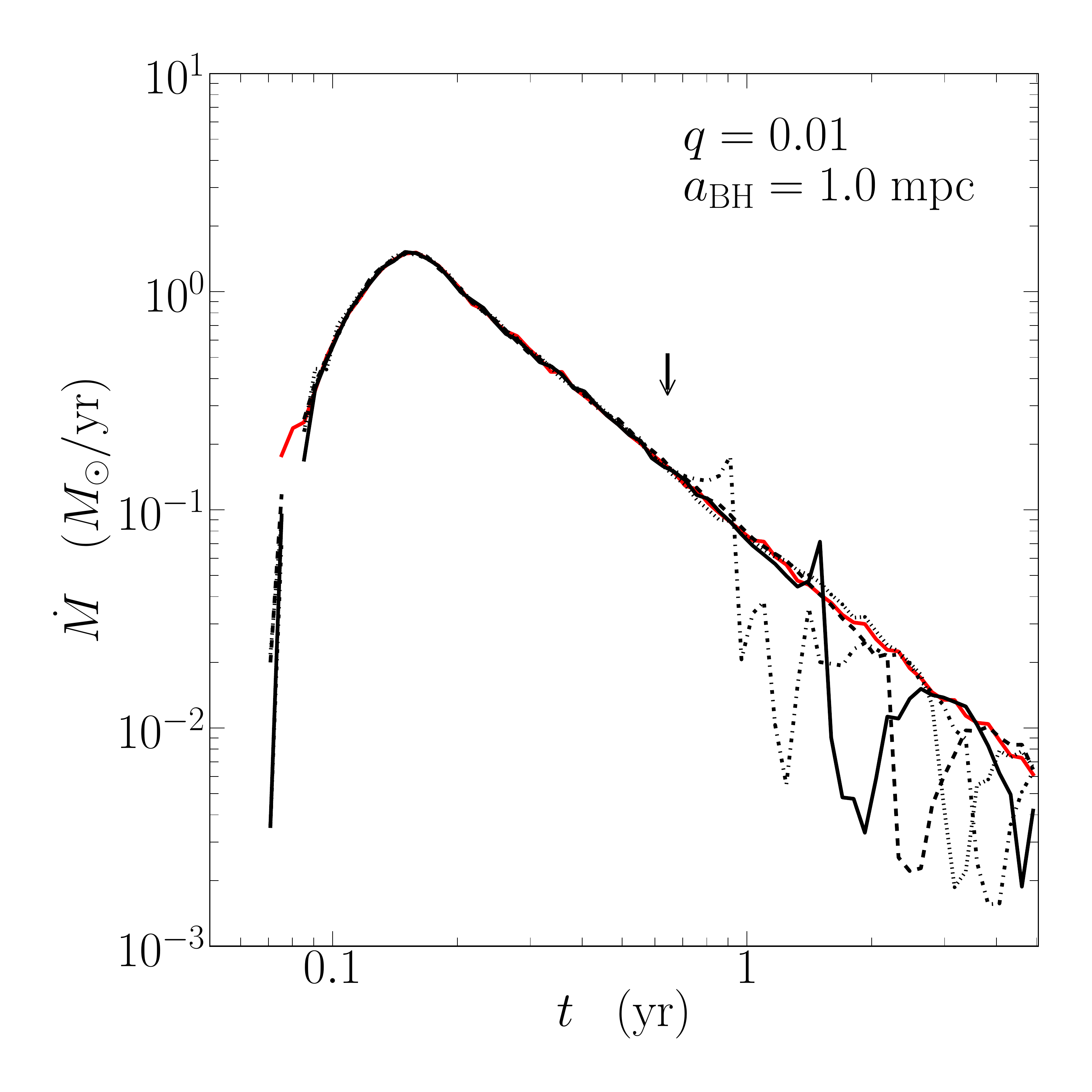}
	\includegraphics[width=0.66\columnwidth]{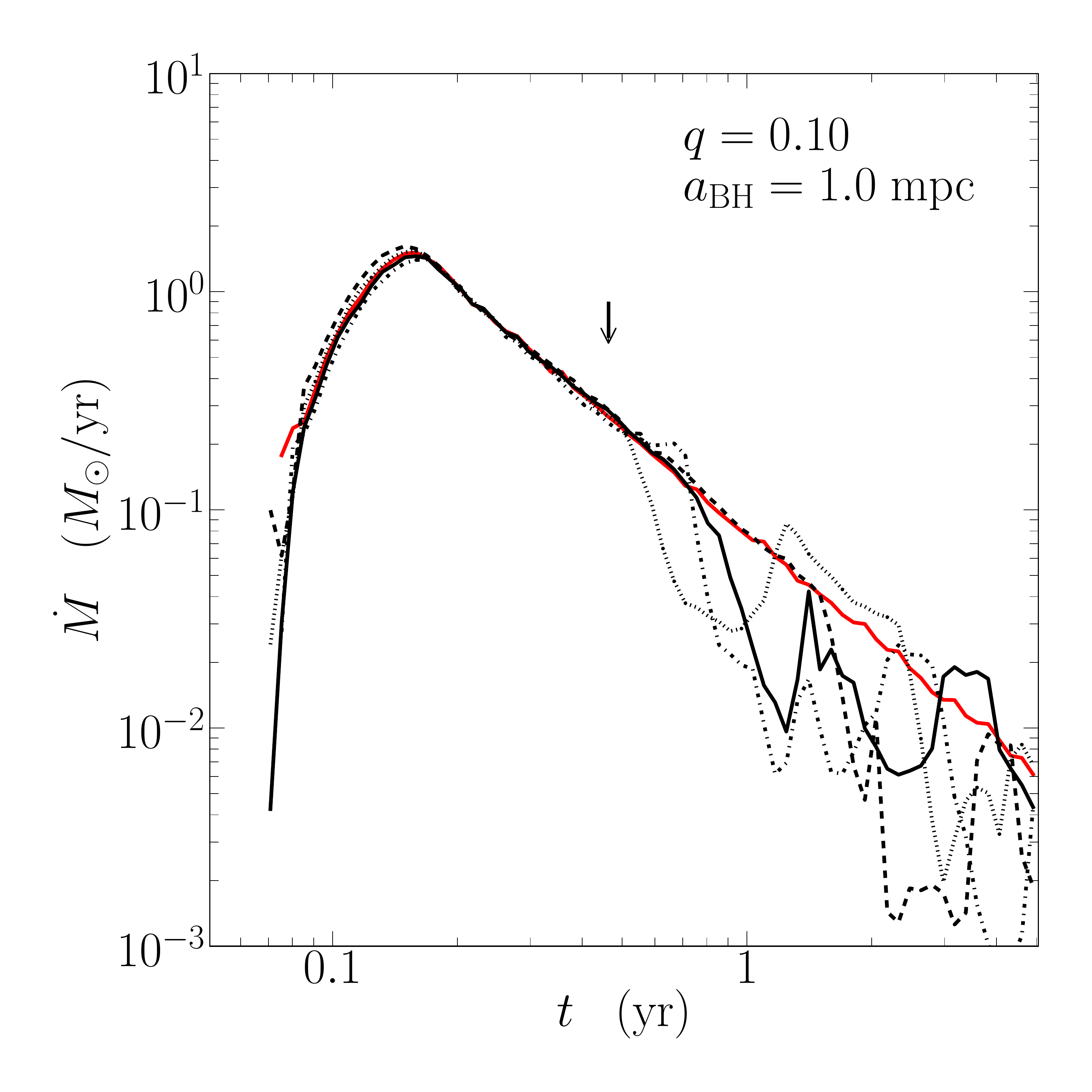}
	\includegraphics[width=0.66\columnwidth]{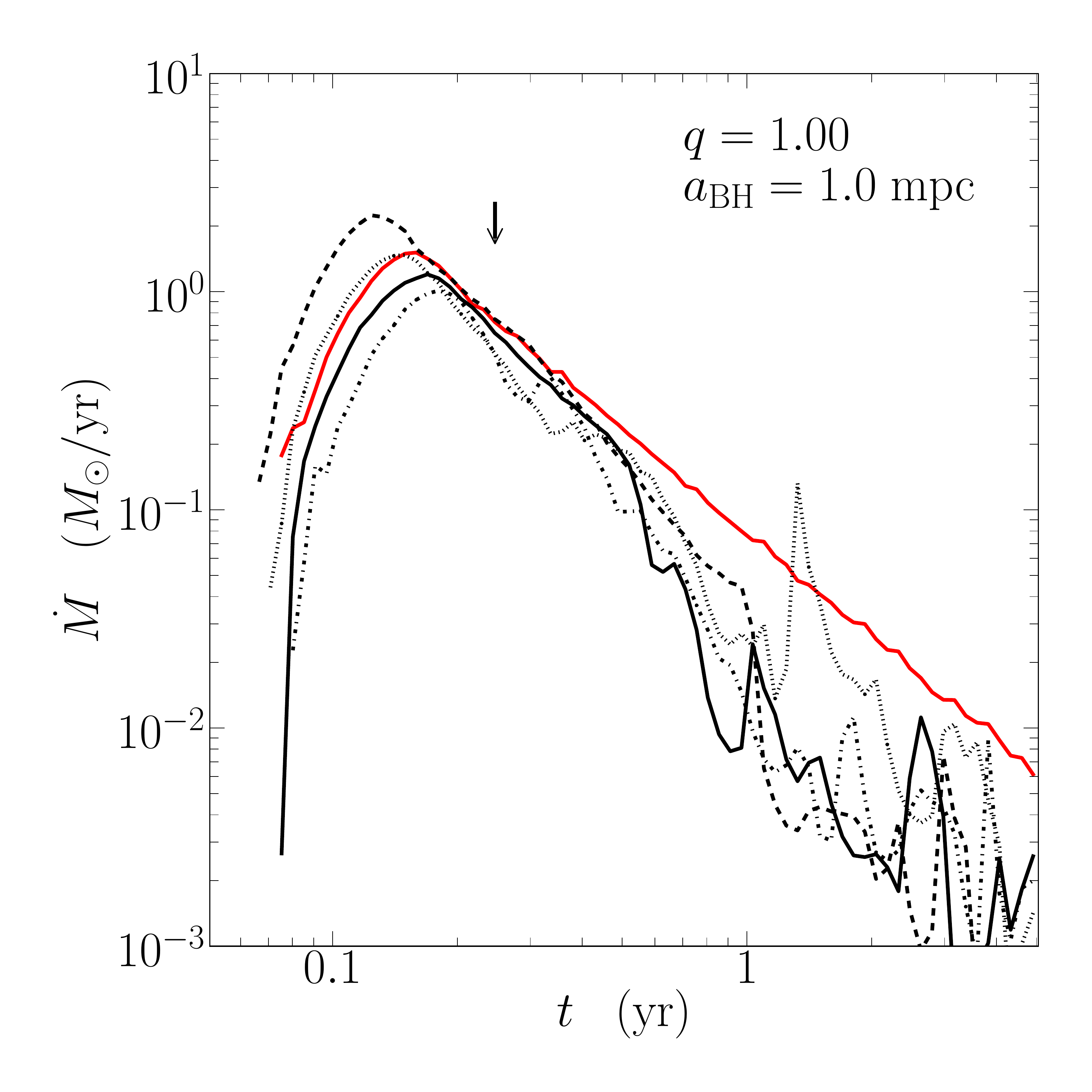}
	\includegraphics[width=0.66\columnwidth]{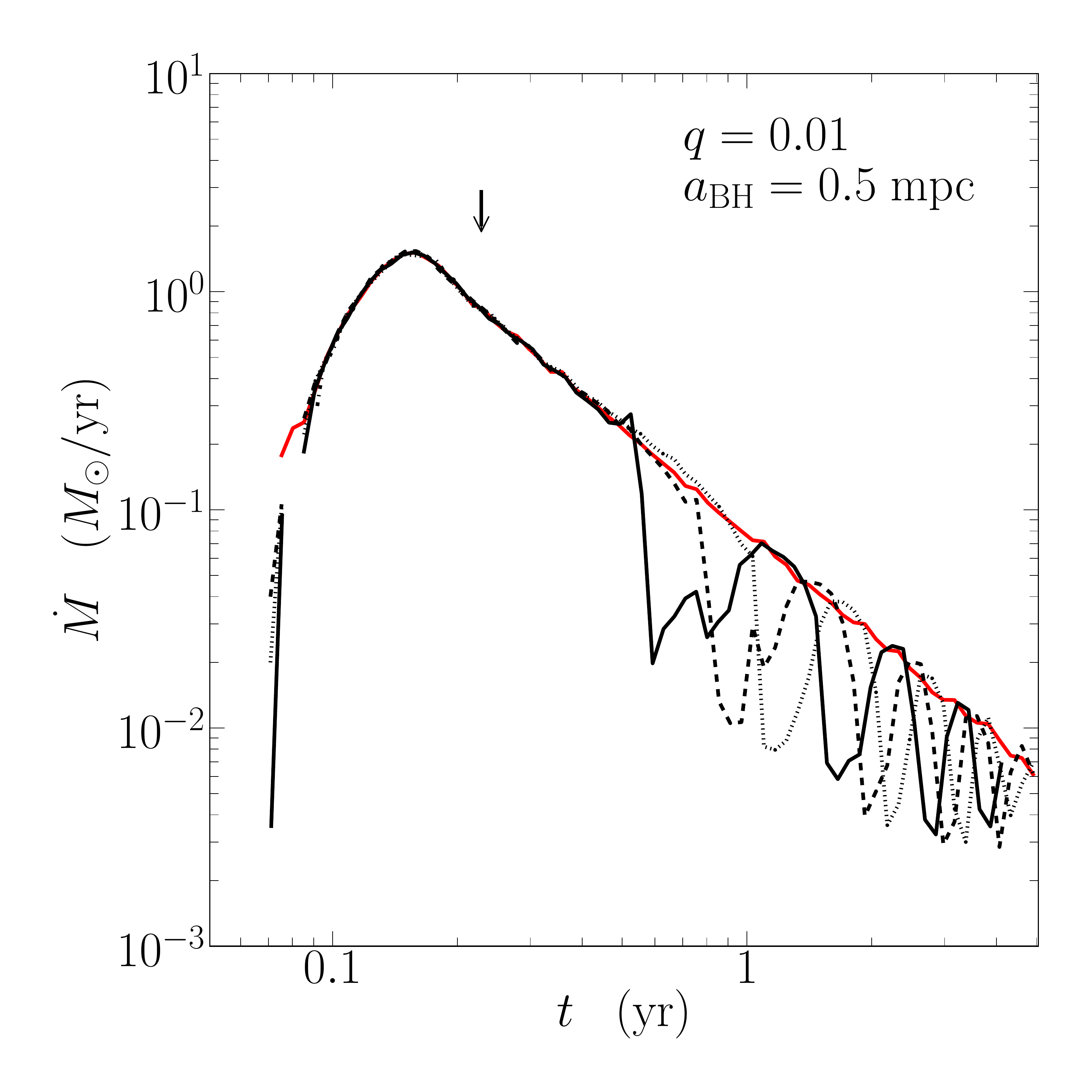}
	\includegraphics[width=0.66\columnwidth]{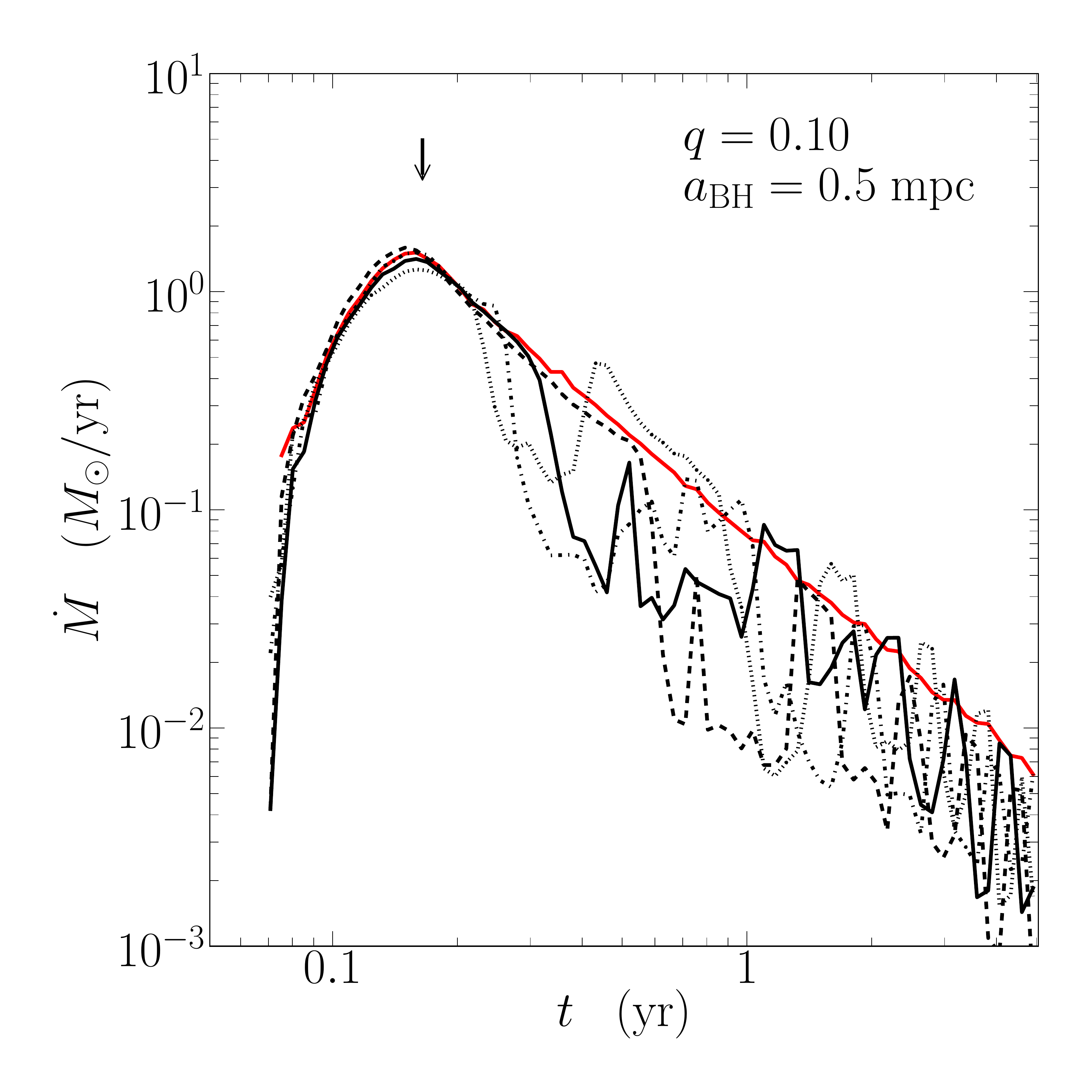}
	\includegraphics[width=0.66\columnwidth]{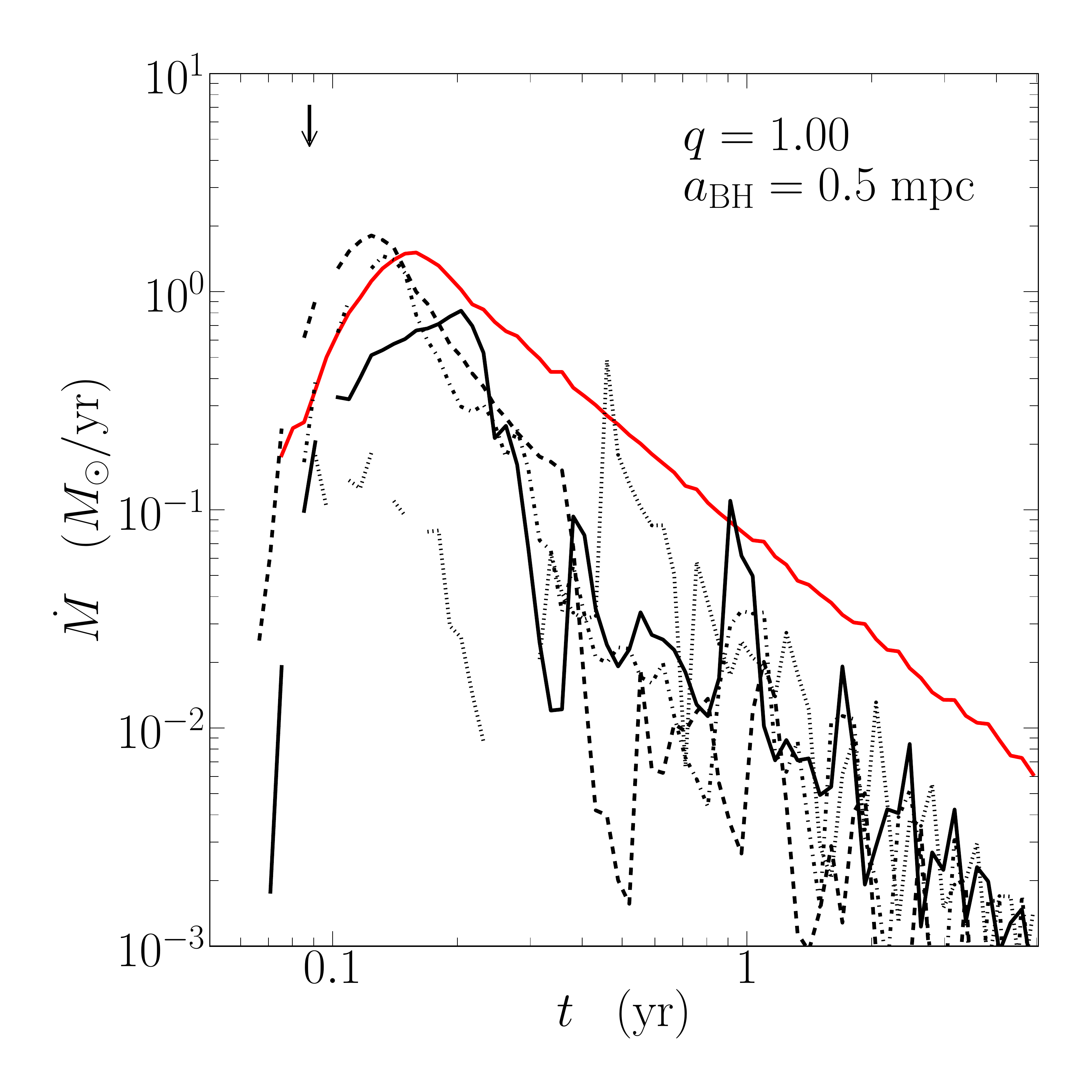}
	\caption{Accretion rate into the primary in function of time in the case of a TDE occurring in the plane. Each panel corresponds to a different set of mass ratio and binary separation according to Figure~\ref{fig::a_minmax} and this is specified in the top-right corner of each graphic. The red curve is the single black hole case. The black curves correspond to the binary case with $\theta=\omega=0^\circ$ and different $\Omega$: $\Omega=0^\circ$ in black full line, $\Omega=90^\circ$ in black long-dashed line, $\Omega=180^\circ$ in black short-dashed line and $\Omega=270^\circ$ in black dot-dashed line. The vertical arrows represent the theoretical truncation time corresponding to each curve.}
	\label{fig::AccRate_plane}
\end{figure*}

Figure~\ref{fig::AccRate_plane} shows the accretion rate onto the primary for each pair of mass ratio and binary separation of Figure~\ref{fig::a_minmax} as indicated on the top-right of each panel, in the case of a disruption in the plane. The red curves are plotted for comparison and corresponds to the single black hole case (SBHC). The black curves are the accretion rate for the binary case for different $\Omega$ and with $\theta=\omega=0^\circ$  ($\Omega=0^\circ$ in full line, $\Omega=90^\circ$ in  long-dashed line, $\Omega=180^\circ$ in  short-dashed line and $\Omega=270^\circ$ in  dot-dashed line). The vertical arrows represent the theoretical truncation time. The simulations are stopped after 5 years. The time at which the luminosity is $1\%$ of the peak luminosity is approximately 3 years.

For each plot (except for $\{q=1; a_{\rm BH}=0.5~\text{mpc}\}$) and each $\Omega$, the accretion rate follows the SBHC in the beginning. Then it undergoes several sharp drops (interruptions) and retrievals of the SBHC. The first interruption arises rarely at the theoretical truncation time and depends mainly on $\omega$. The periodicity of the drops is not always present. For instance for $\{q=1; a_{\rm BH}=1~\text{mpc}\}$, after the first drop, the accretion rate becomes chaotic.

When we have a periodicity, the period of the retrievals is approximately the orbital period of the black holes. Each drop corresponds to the passage of the secondary into the stream of in-falling debris. During this passage, the secondary disturbs a part of the stream. This part will not return normally at the pericenter, or eventually at a delayed time, and then the accretion rate undergoes an interruption. 

In conclusion, when the disruption is in the plane, the process that leads to the interruption of the accretion rate results from the close passages of the secondary near the stream of debris. That is why there is a big dependency of the time of first interruption on the initial $\Omega$, i.e. on the azimuthal position of the stream. For instance, for $\Omega=90^\circ$, the stream is sent on the opposite side of the initial position of the secondary and the first interruption occurs at a later time than for $\Omega=270^\circ$ (see Figure~\ref{fig::AccRate_plane}). This result confirms the one of \citet{Ricarte2015}.

\citet{Coughlin15} and \citet{Coughlin16b} have shown that the stream of debris after a tidal disruption event by a single black hole can become gravitationally unstable and form clumps, which may add a level of variability in the fallback rate. While this is not the main focus of this paper, we confirm that similar clumps also appear in the case of disruption by binary black holes, implying that the additional tidal shear due to the binary potential is not sufficient to prevent self-gravitating clump formation.

\subsubsection{Disruption perpendicular to the plane}

\begin{figure*}
	\centering
	\includegraphics[width=0.66\columnwidth]{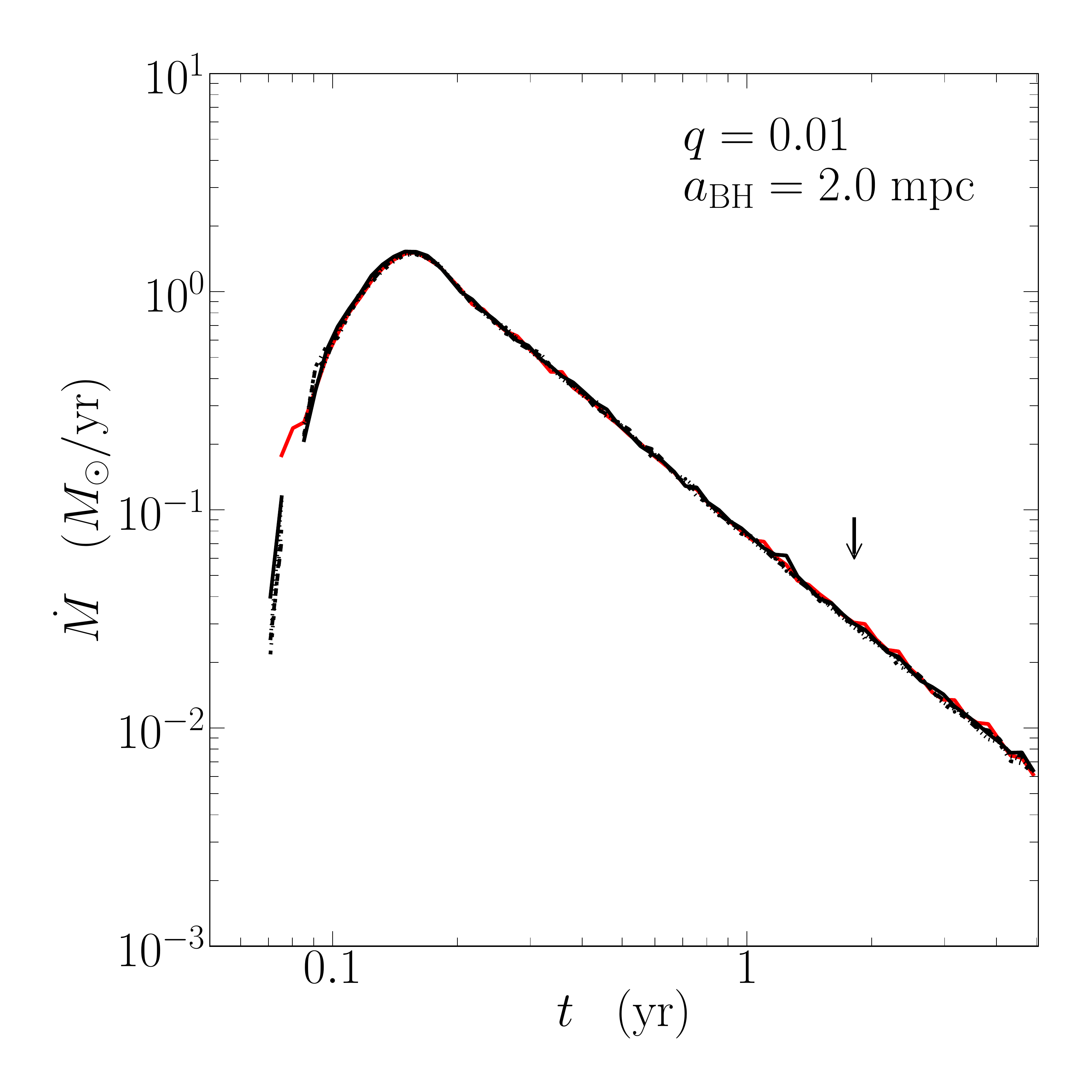}
	\includegraphics[width=0.66\columnwidth]{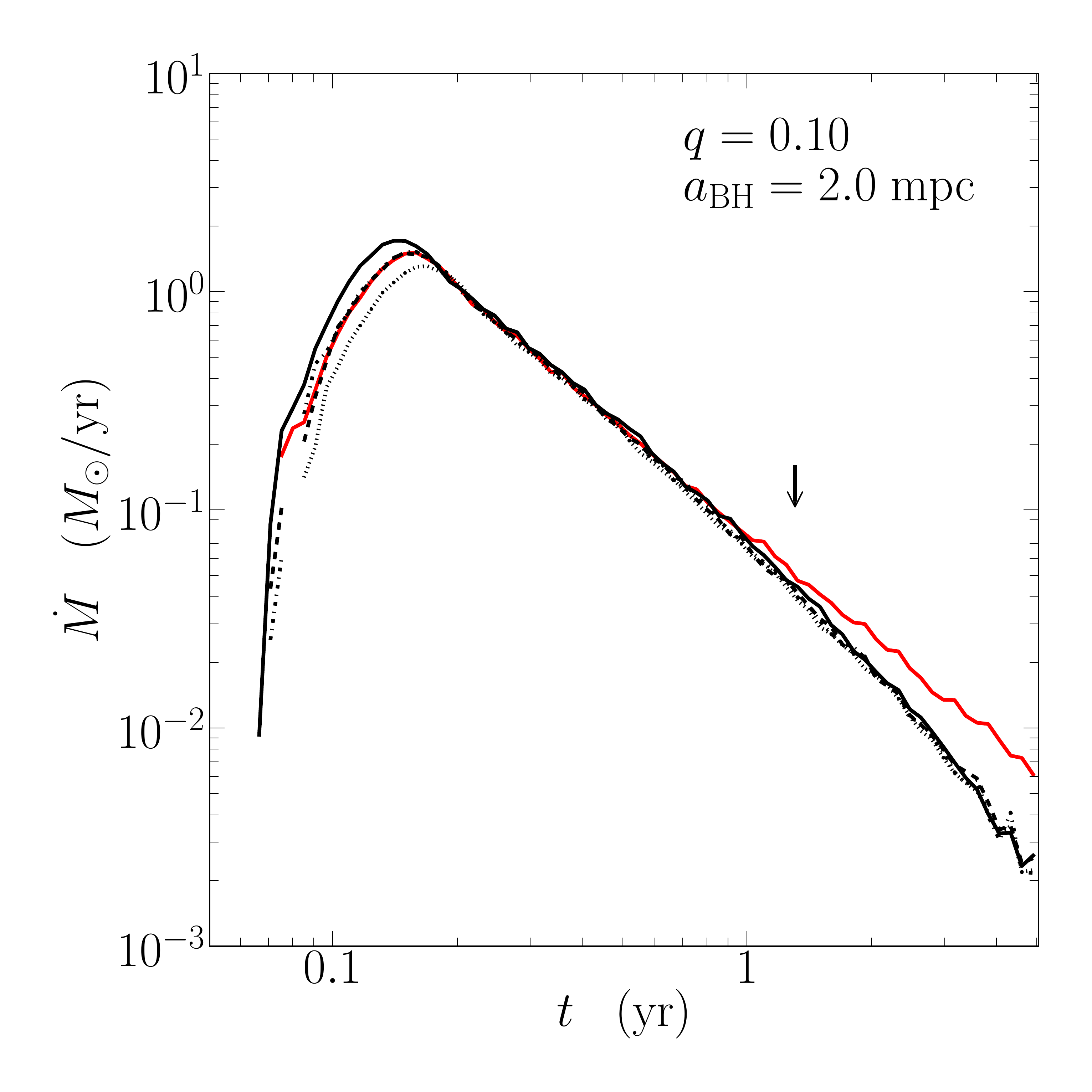}
	\includegraphics[width=0.66\columnwidth]{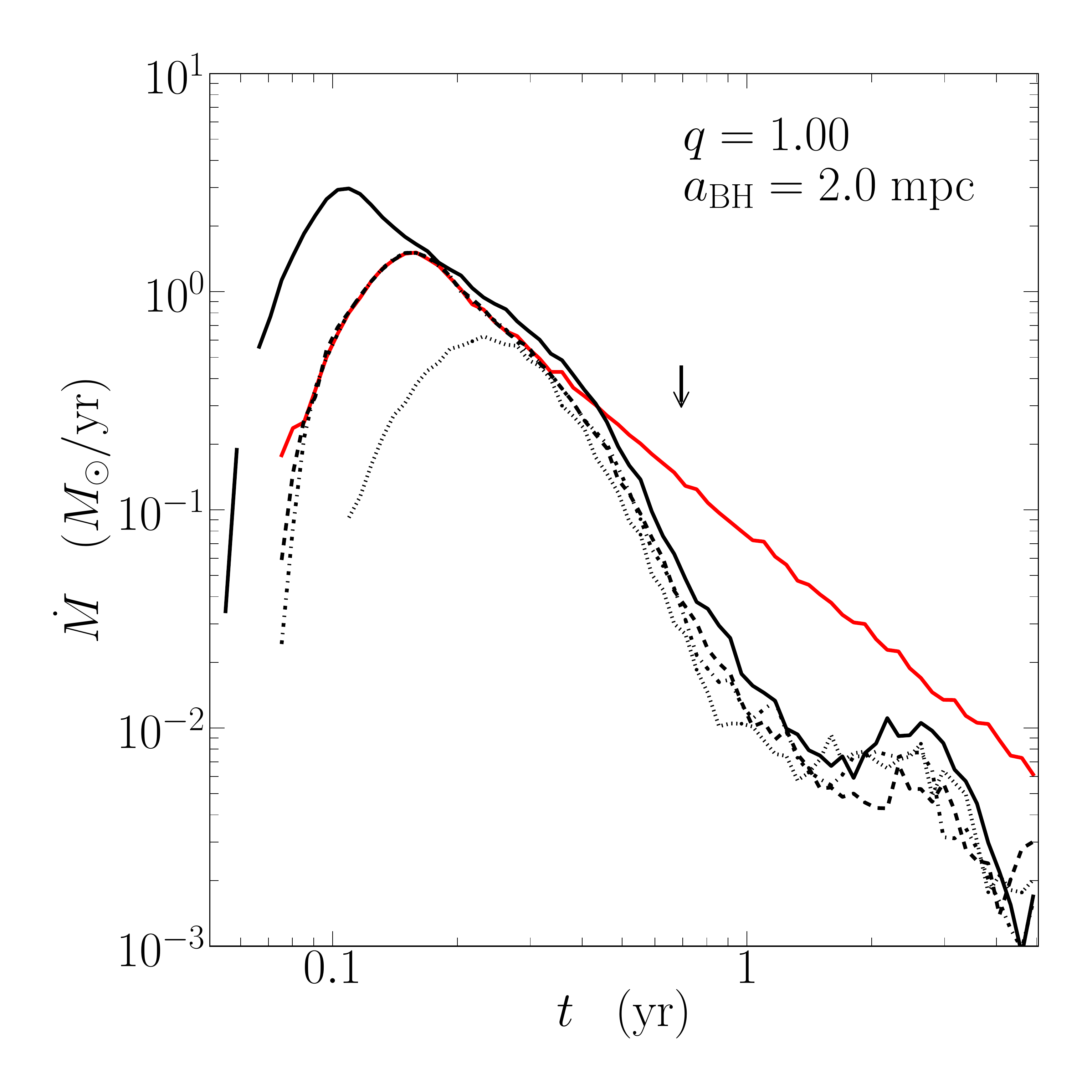}
	\includegraphics[width=0.66\columnwidth]{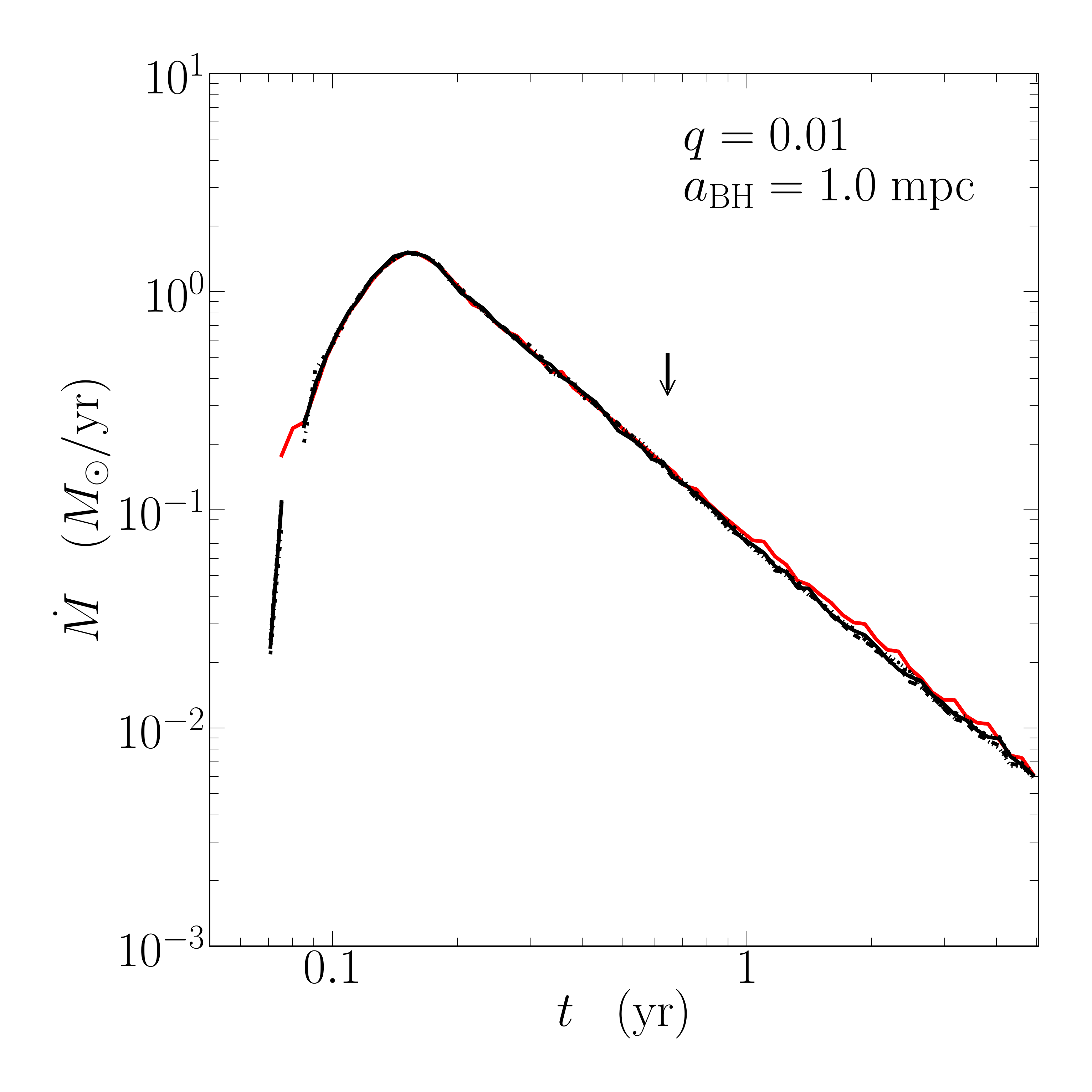}
	\includegraphics[width=0.66\columnwidth]{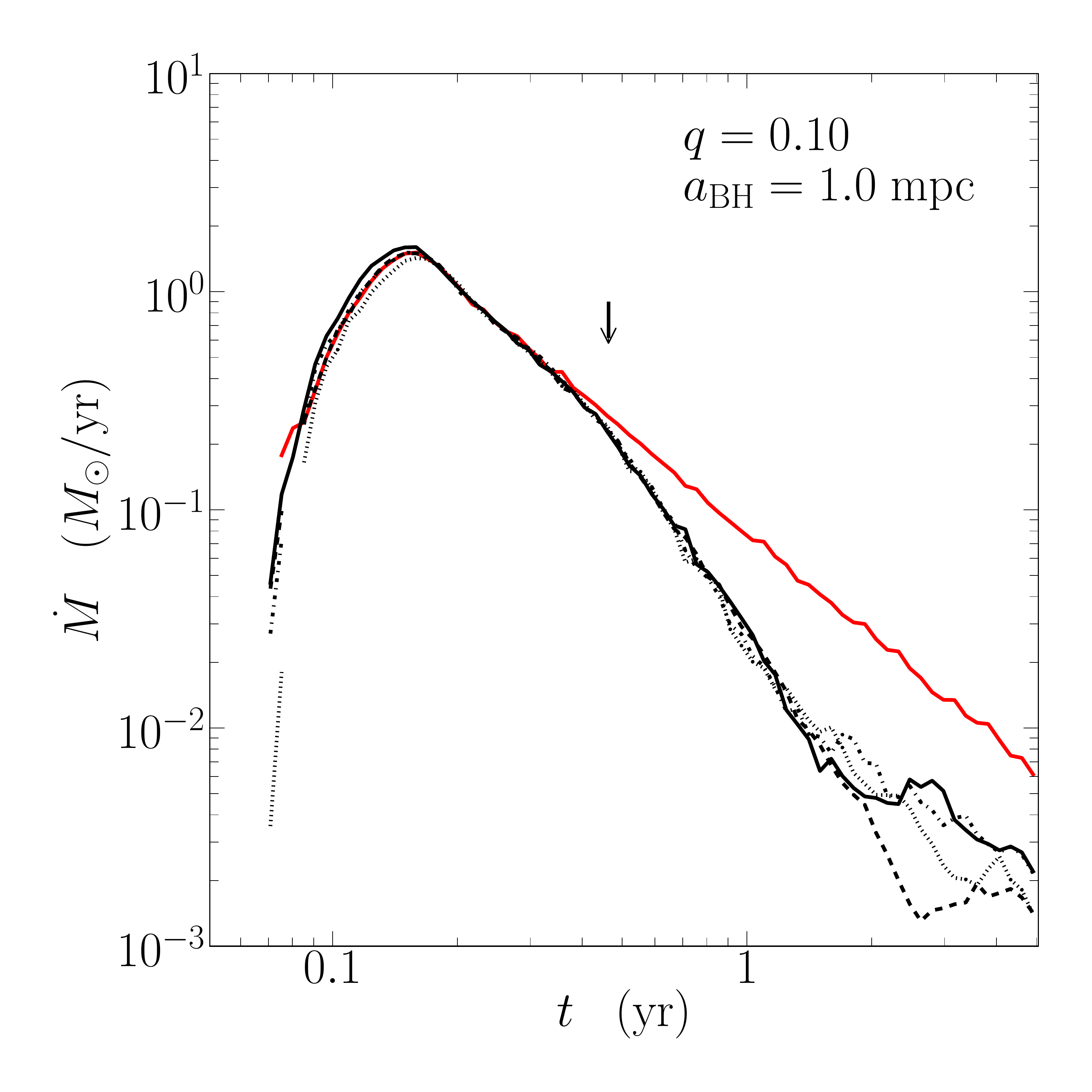}
	\includegraphics[width=0.66\columnwidth]{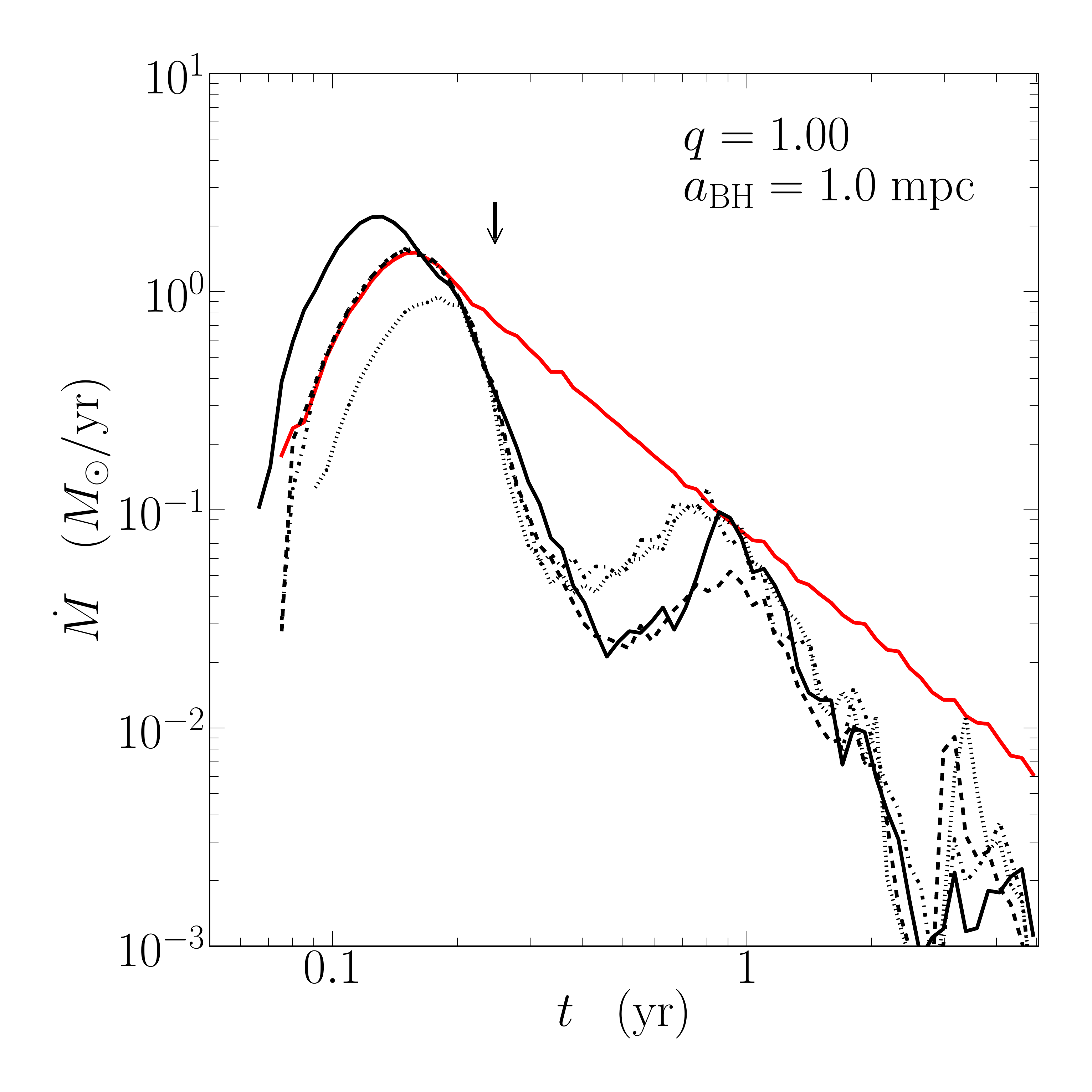}
	\includegraphics[width=0.66\columnwidth]{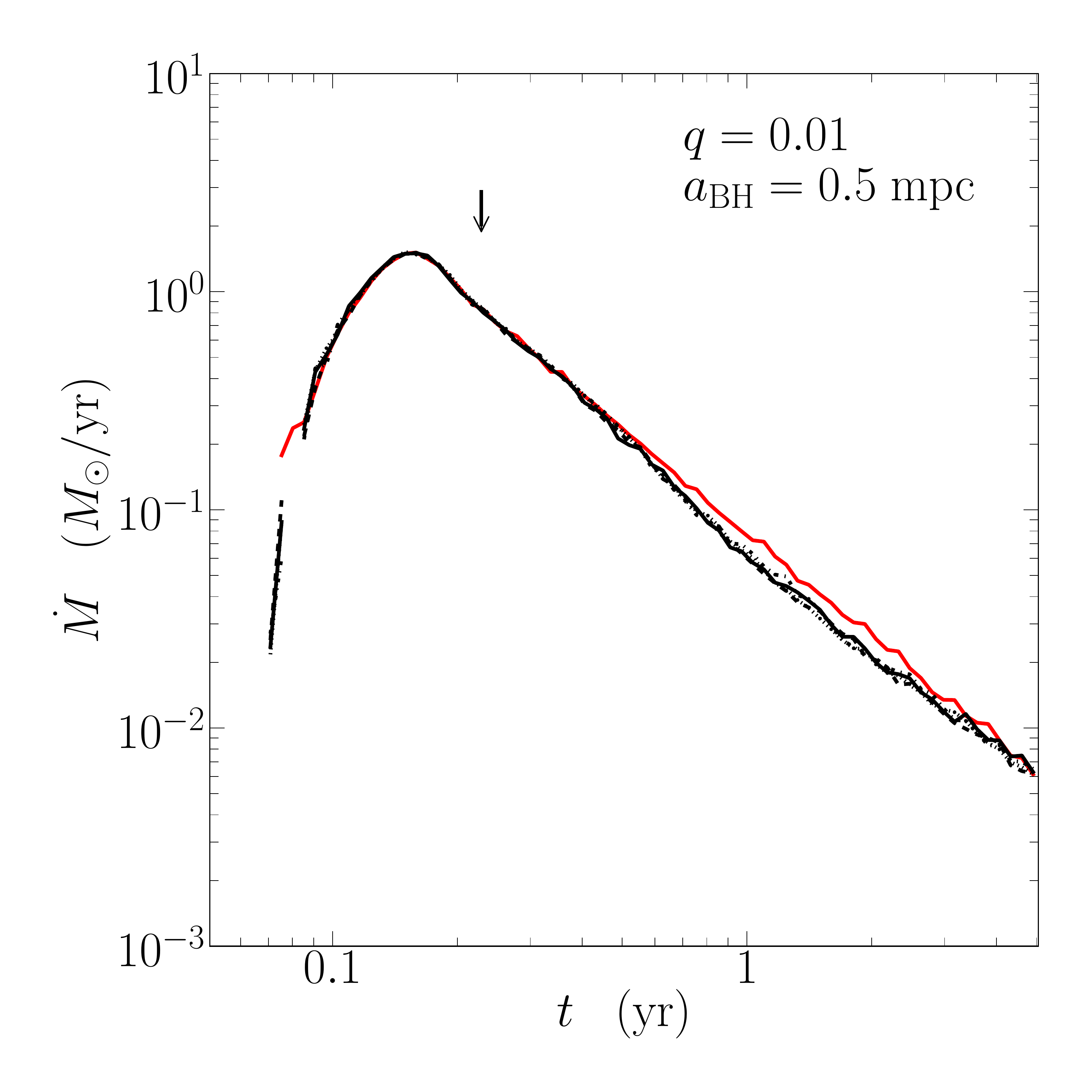}
	\includegraphics[width=0.66\columnwidth]{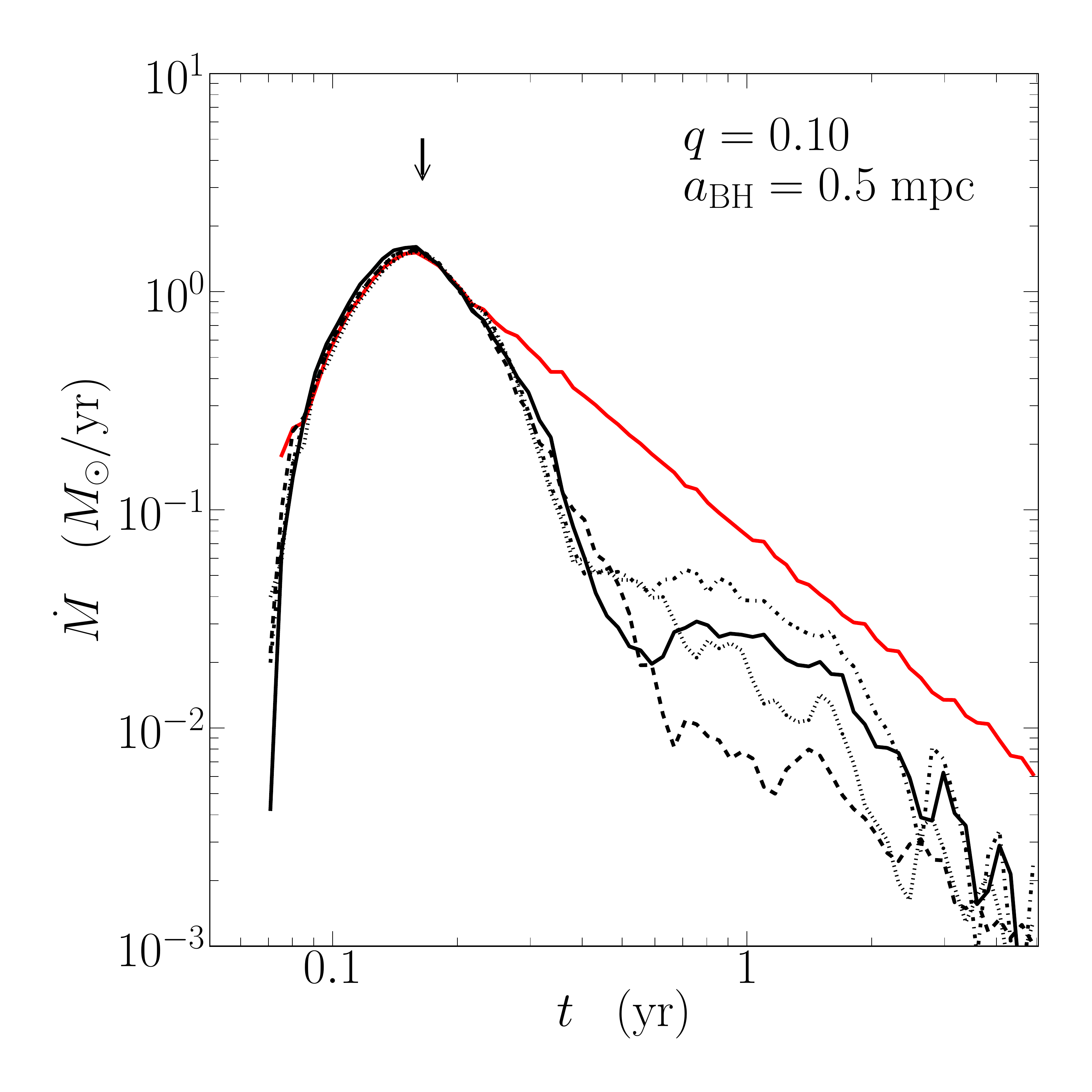}
	\includegraphics[width=0.66\columnwidth]{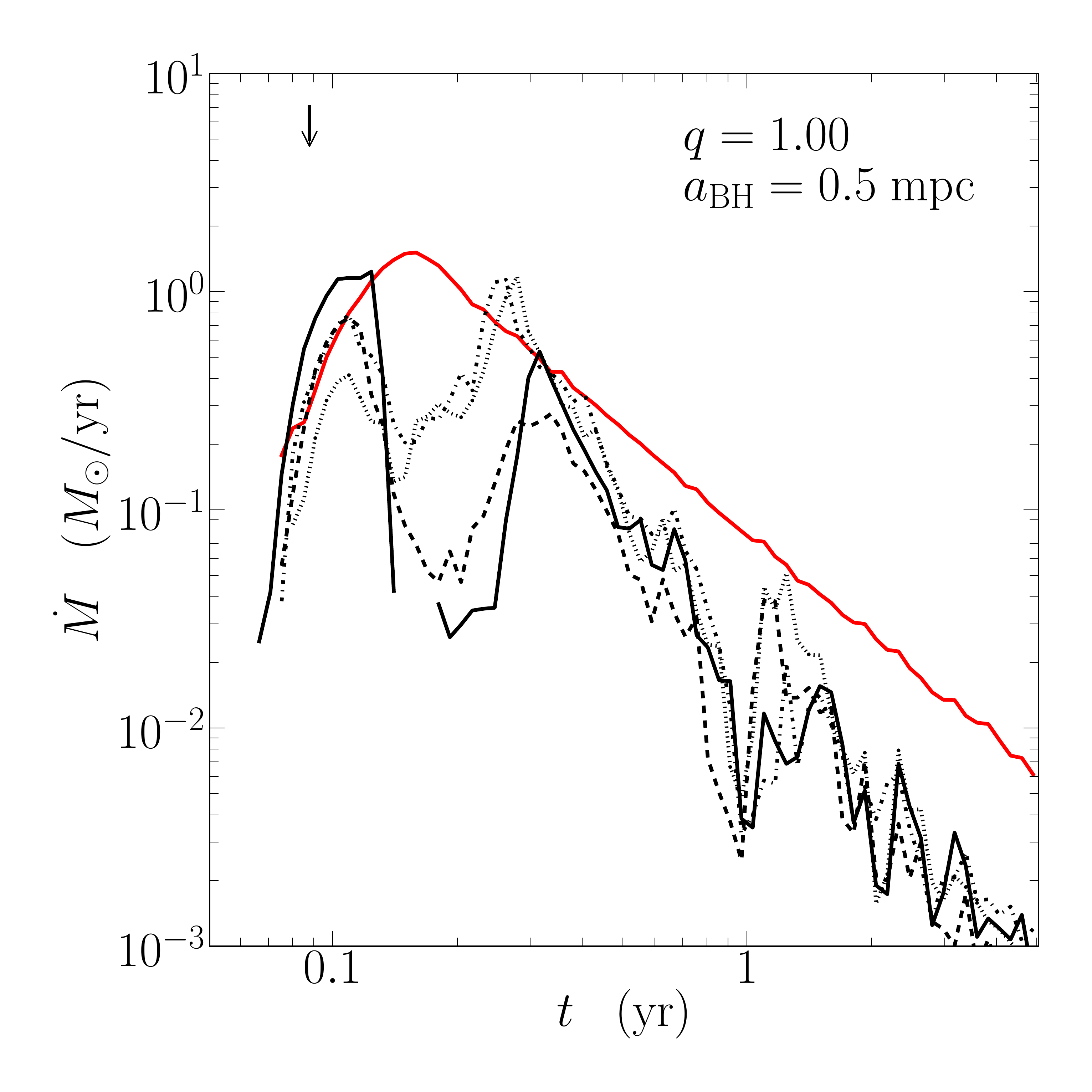}
	\caption{Accretion rate into the primary in function of time in the case of a TDE occurring perpendicular to the plane. Each panel corresponds to a different set of mass ratio and binary separation according to Figure~\ref{fig::a_minmax} and this is specified in the top-right corner of each graphic. The red curve is the single black hole case. The black curves correspond to the binary case with $\theta=\omega=90^\circ$ and different $\Omega$: $\Omega=0^\circ$ in black full line, $\Omega=90^\circ$ in black long-dashed line, $\Omega=180^\circ$ in black short-dashed line and $\Omega=270^\circ$ in black dot-dashed line. The vertical arrows represent the theoretical truncation time corresponding to each curve.}
	\label{fig::AccRate_perp}
\end{figure*}

Figure~\ref{fig::AccRate_perp} shows the accretion rate onto the primary for each pair of mass ratio and binary separation of Figure~\ref{fig::a_minmax} as indicated on the top-right of each panel, in the case of a disruption perpendicular to the plane. The red curves are plotted for comparison and corresponds to the SBHC. The black curves are the accretion rate for the binary case for different $\Omega$ and with $\theta=\omega=90^\circ$  ($\Omega=0^\circ$ in  full line, $\Omega=90^\circ$ in  long-dashed line, $\Omega=180^\circ$ in short-dashed line and $\Omega=270^\circ$ in dot-dashed line). As before, the vertical arrows represent the theoretical truncation time and the simulations are stopped after 5 years. 

For $q=0.1$, $\{q=1; a_{\rm BH}=1~\text{mpc}\}$ and $\{q=1; a_{\rm BH}=2~\text{mpc}\}$ we observe, as in Section~\ref{sec::dis_plane}, an initial decline, following the single black hole case and then a first interruption. However, this interruption is not always followed by a retrieval of the SBHC and is largely smoother than the ones in the disruptions in the plane. Moreover the time of the first drop does not depend on $\Omega$ and corresponds to the theoretical truncation time. We have a totally different behavior and process that leads to the interruptions of the SBHC in the case where the disruption is perpendicular to the plane.

This difference can be explained by the way the secondary interacts with the debris. When the disruption is in, or close to, the plane, the secondary has periodic very close passages near the in-falling stream. This results in violent periodic disturbances of the stream, and also very net interruptions of the accretion rate. However, when the disruption is out of the plane, the stream is mainly located in the $z$-axis (the location does not depend on $\Omega$) and the distance between one part of the stream and the secondary is constant. Thus the perturbation created by this black hole on the stream is not periodic and is of the same order all the time. This leads to a smooth interruption of the accretion rate. Moreover, because the stream is mainly located in the $z$-axis, $\Omega$ has no major influences on the relative position of this stream with respect to the secondary and that is why there is no dependency of the time of first interruption in $\Omega$.

One should note that this behavior is not present in the case $q=0.01$, where the accretion rate follows the SBHC for each $\Omega$ and $a_{\rm BH}$. We can conclude that the process leading to the perturbations of the stream in the case of a perpendicular disruption is not efficient for small $q$. Then, for $q=0.01$, only a close encounter of the secondary with the stream, i.e. disruption close to the plane, can perturb this stream and create visible interruptions (see Figure~\ref{fig::AccRate_plane}, left panels).

\subsubsection{Boundary between the two behaviors}

We observed interruptions of the accretion rate due to the secondary. These interruptions behave differently depending on the initial orientation of the orbit of the star.  If the initial orbit is in the plane of the SMBHB, sharp and periodic interruptions occur. On the contrary, if the initial orbit is perpendicular to the plane, a first very smooth interruption occurs, which is eventually, but not always, followed by a retrieval of the SBHC before the maximum observational time. Then a new question appears: what is the critical inclination $\theta_{\rm cr}$ separating those two behaviors?

\begin{figure}
	\centering
	\includegraphics[width=\columnwidth]{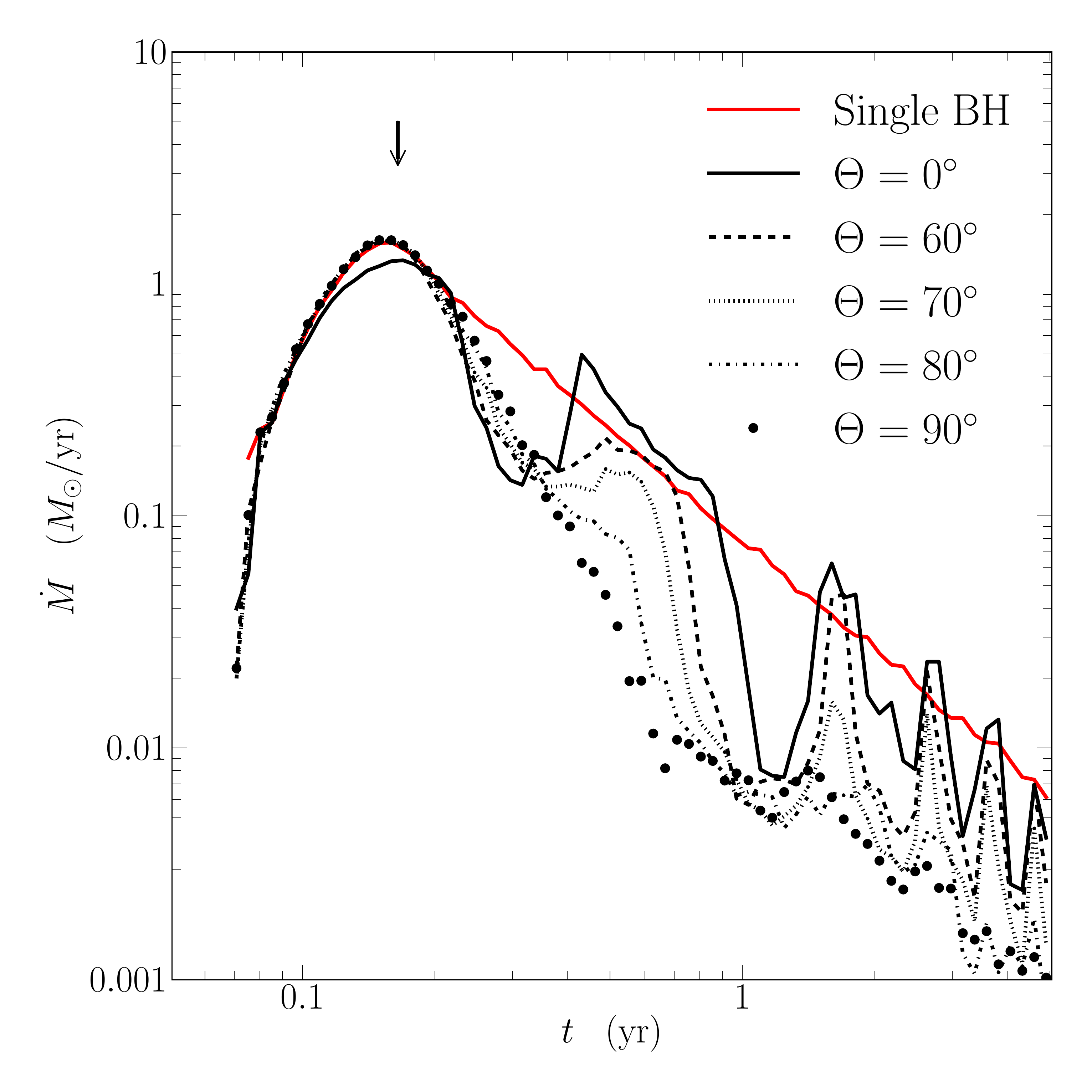}
	\caption{Accretion rate into the primary for $q=0.1$, $a_{\rm BH}=0.5~\text{mpc}$ and $\omega=\Omega=90^\circ$. The red curve is the single black hole case. The black curves represent different inclinations $\theta$. The vertical arrow represents the theoretical truncation time corresponding to each curve.}
	\label{fig::AccRate_90_90_big_omega}
\end{figure}

Figure~\ref{fig::AccRate_90_90_big_omega} compares the accretion rate for different inclinations $\theta \in \{0^\circ,60^\circ,70^\circ,80^\circ,90^\circ\}$ (black curves) with $q=0.1$, $a_{\rm BH}=0.5~\text{mpc}$ and $\omega=\Omega=90^\circ$. The red curve is the SBHC and the vertical arrow represents the theoretical truncation time. We see that when the inclination goes from $0^\circ$ to $90^\circ$, the accretion rate evolves smoothly from the periodic behavior to the smoother behavior. Even if we cannot define a precise critical inclination, we see that with $\theta=60^\circ$ there are still periodic retrievals of the SBHC. This only begins to vanish from $\theta=70^\circ$, and has totally vanished for $\theta=80^\circ$.  We can reasonably say that the critical inclination is $\theta_{\rm cr} \sim 70^\circ$.


\subsection{Influence of the binary separation}
\label{sec::influence}

The influence of the binary separation on the accretion rate was not discussed by \citet{Liu2009} and \citet{Ricarte2015}.

Firstly we analyse the case of a disruption in the plane. If we refer to  Figure~\ref{fig::AccRate_plane}, we see that the first interruption occurs always before 3 years (maximum observational time we chose). The exception is the case $\{q=0.01; a_{\rm BH}=2~\text{mpc}$\} where the first interruption occurs before 3 years only for $\Omega=270^\circ$. In this case, the detection of the interruption, and also of the secondary, would be difficult. This is indeed predicted by Equation~\eqref{eq::a_minmax} since the corresponding point in the Figure~\ref{fig::a_minmax} is above the blue area. We can conclude that the upper limit fixed by Equation~\eqref{eq::a_minmax} is valid for a disruption in the plane. If the disruption is perpendicular to the plane (Figure~\ref{fig::AccRate_perp}), an interruption occurs only for $q=0.1$ and $q=1$. Thus Equation~\eqref{eq::a_minmax} overpredicts the effects of the secondary for the smallest mass ratios.

Concerning the lower limit, the same conclusion arises. By analyzing the case $\{q=1; a_{\rm BH}=0.5~\text{mpc}\}$ in which the theoretical truncation time is smaller than $t_{\rm min}$, we see that at no moment the SBHC is followed and the accretion rate is chaotic. So not only we will not be able to detect the secondary, but also it will not be possible to determine the lightcurve to be resulting from a TDE, at least with the method using the power law. Thus the case $\{q=1; a_{\rm BH}=0.5~\text{mpc}\}$ depicts the fact that for very small binary separations, the most bound debris are disturbed by the secondary before they return to the pericenter. The usual description of TDEs, made by the power law, does not holds anymore. The lower limit of Equation~\eqref{eq::a_minmax} accounts well for this phenomenon.

In conclusion, the limits set by Equation~\eqref{eq::a_minmax} are a good restriction of the binary separation needed for detecting the secondary if the disruption is in the plane. In the case it is perpendicular to the plane, the limits only hold for $q \gtrsim 0.1$.

\section{Accuracy of fallback rates}
\label{sec::limits}

Care should be taken about the method used to compute the fallback rate from the numerical simulations.

\citet{Liu2009}, \citet{Ricarte2015} and \citet{Coughlin2016} computed the accretion rate using the accretion radius of the primary. A major drawback of this method is that it depends mainly on the value of this radius, as quoted in \citet{Ricarte2015}, and clearly shown also in \citet{Coughlin2016}, where they evaluate two different fallback rates using different choices for the accretion radius. In this article we chose to use a different method, which makes use of Equation~\eqref{eq::fallback_rate}.

For each time $T$ we compute the mass distribution $\text{d}M/\text{d}E_1$, where $E_1$ is the orbital energy of a particle with respect to the primary. $\text{d}M/\text{d}E_1$ depends on $E_1$ and $T$. Following the same methods as \citet{Rees1988}, Kepler's third law gives us $E_1$ as a function of $T$. Then we obtain $\text{d}M/\text{d}E_1$ as a function only of $T$, and finally with Equation~\eqref{eq::fallback_rate} we get $\text{d}M/\text{d}T$ at the time $T$. We repeat these steps for each time $T$ by recomputing the mass distribution at these times.

 For simplicity, we call our method (using the mass distribution) ``method 1", and the method using the accretion radius ``method 2".

\begin{figure}
	\centering
	\includegraphics[width=\columnwidth]{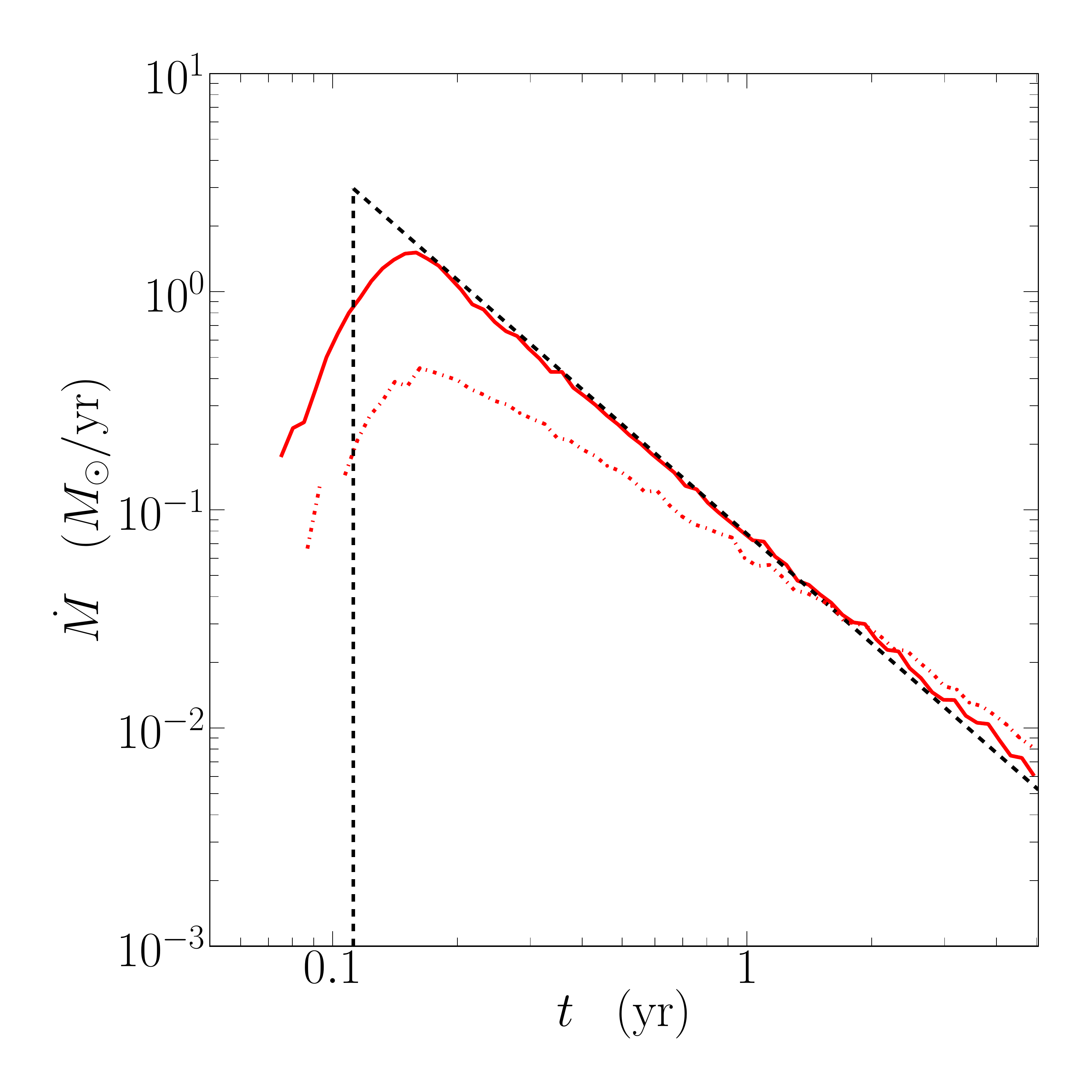}
	\caption{Accretion rate onto the primary as a function of time in the case of a single black hole of mass $M_{\rm h}=10^6\msol$. The black dashed line represents the theoretical power law of Equation~\eqref{eq::dMdT_uni}, the red full line is the accretion rate calculated with method 1 and the red dot-dashed line with method 2. We see that method 1 fits well the power law at late times, while method 2 does not fit at any time the power law.}
	\label{fig::Comparison_method}
\end{figure}

Figure~\ref{fig::Comparison_method} compares the accretion rate as a function of time calculated with the two methods (method 1 in red full line and method 2 in red dot-dashed line) with the theoretical power law of Equation~\eqref{eq::dMdT_uni} (dashed black line)  for the case of a single black hole of mass $M_{\rm h} = 10^6\msol$. Only  method 1 fits the expected power law well. Because of the polytropic shape of the star this is only true at late times as shown by \citet{Lodato2009}. The peak luminosity of method 2 is significantly below the theoretical peak and the $t^{-5/3}$ behavior is never reached, even at late times. This confirms that our method is well appropriate for a single black hole.

\begin{figure*}
	\centering
	\includegraphics[width=\columnwidth]{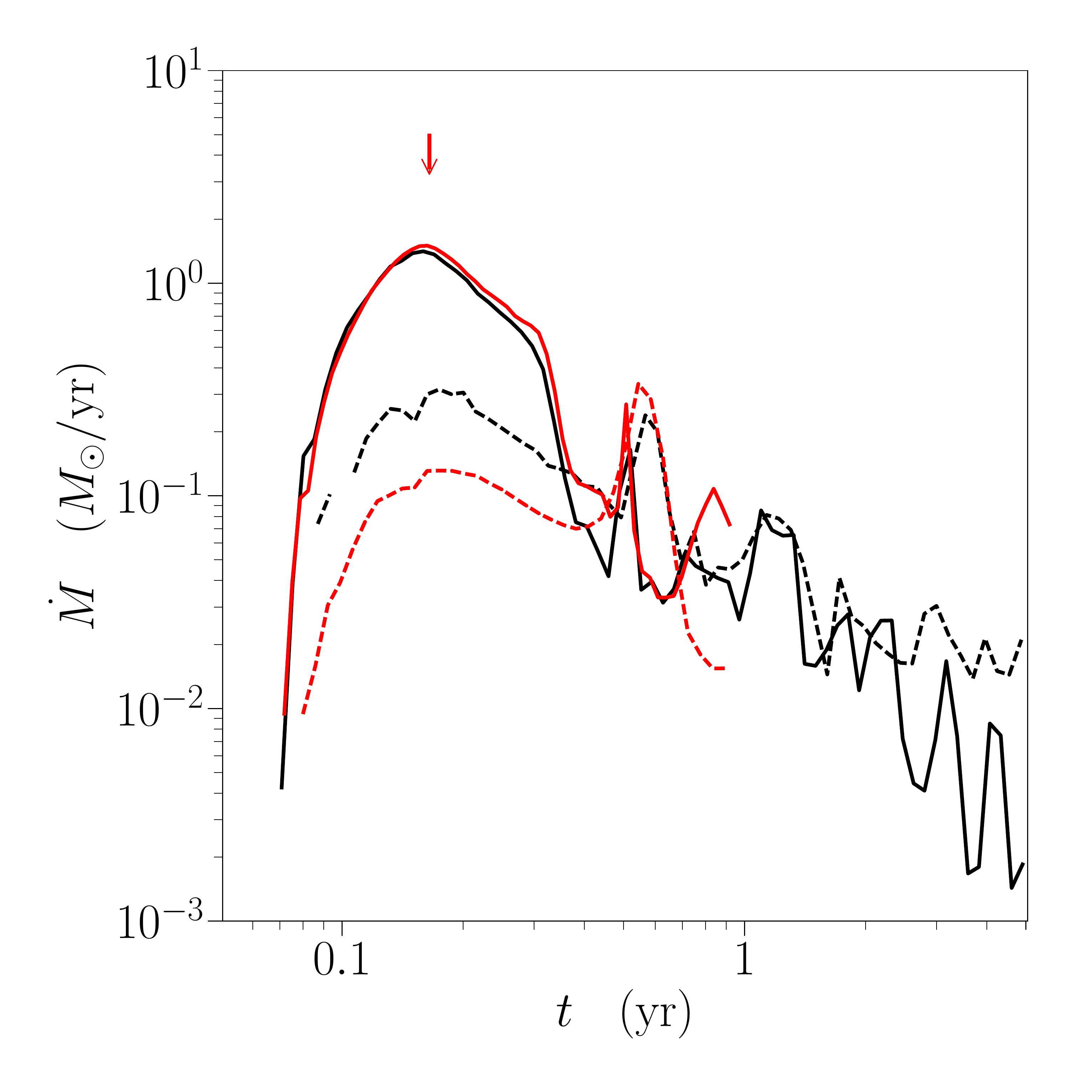}
    \includegraphics[width=\columnwidth]{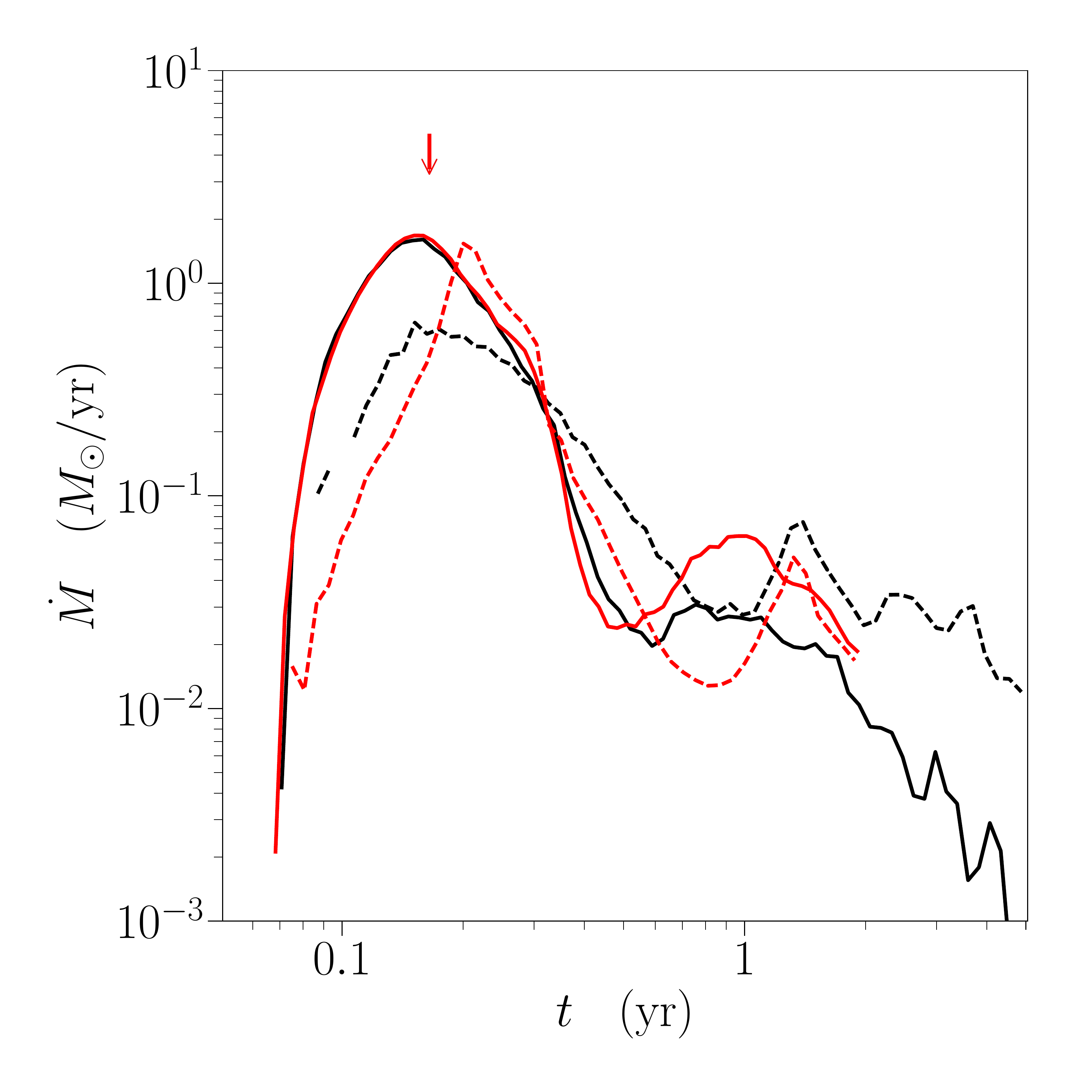}
	\caption{Comparison between the fallback rate computed by using our method based on the energy distribution of the debris (method 1, solid line) and by using an accretion radius (taken to be $0.8 R_{\rm p}$, dashed line). The two panels show the case with $10^5$ particles, a mass ratio $q=0.1$, a separation $a_{\rm BH}=0.5$ mpc, for an encounter in the orbital plane of the binary ($\theta=\omega=\Omega=0^\circ$, left panel) and perpendicular to it ($\theta=\omega=90^\circ$ and $\Omega=0^\circ$, right panel). The red lines refer to the same simulations, but at the higher resolution of $10^6$ particles, and again the solid lines refer to method 1 and the dashed lines to method 2.}
	\label{fig::Comparison_method_binary}
\end{figure*}

Obviously, since the gravitational potential in the case of a binary black hole system is not Keplerian, we do not expect the debris to fall back exactly at the rate predicted by our method. Still, our method can be regarded as a way to measure the changes in the debris specific energies due to the presence of the binary companion. If the distribution is unperturbed (negligible modifications to the standard $t^{-5/3}$ regime), the binary does not affect the disruption, while significant changes to the $t^{-5/3}$ decline indicate that debris that are expected to fall back at a given time will be strongly influenced by the binary potential. 

In order to estimate the uncertainty in the fallback rate associated with our choice, we plot in Figure \ref{fig::Comparison_method_binary} the fallback rates estimated with our method (black solid line) and using an accretion radius, taken to be $0.8 R_{\rm p}$ (black dashed line). The two panels refer in particular to the case with a mass ratio $q=0.1$, a separation $a_{\rm BH}=0.5$ mpc, for an encounter in the orbital plane of the binary ($\theta=\omega=\Omega=0^\circ$, left panel) and perpendicular to it ($\theta=\omega=90^\circ$ and $\Omega=0^\circ$, right panel). The major difference in the two calculations is that the initial fallback rate is much smaller when using method 2, in line with our findings for a single black hole. In this respect, we can therefore state that at least initially our method represents more faithfully the fallback of the debris. At late times, when the interaction with the secondary becomes important and the fallback is interrupted, we note that the two methods give comparable results, although the exact details clearly change somewhat. In particular, the main feature that for an encounter in the binary plane the fallback undergoes a series of quasi periodic interruptions while for a perpendicular encounter there is a smoother reduction in the fallback rate is evident when using both methods. Fig. \ref{fig::Comparison_method_binary} also shows, with red lines, the results of a convergence test, where the fallback rates with the two methods have been evaluated based on a simulation with 10 times more particles than our standard one, i.e. with $10^6$ particles. Again, the solid lines refer to the fallback computed based on the energy distribution of the debris (method 1), while the dashed lines refer to that computed based on the sink radius (method 2). It can be immediately seen that method 1 is much more robust and less sensitive to resolution than the accretion radius method in order to estimate the fallback rate. Indeed, while with method 1 the fallback rate is only mildly modified at high resolution, we can see that using the accretion radius changes significantly the estimated rate already from the initial phases, resulting in a suppression of the fallback rate as resolution is increased. These results give us confidence of the validity of the results presented in the Sections above.

\addcontentsline{toc}{section}{Conclusion}
\section{ConclusionS}

We have performed numerical simulations of the tidal disruption event (TDE) of a star by a supermassive black hole binary (SMBHB) using Smooth Particle Hydrodynamic. The main results of this work are as follow:

\begin{enumerate}

\item The fallback rate to pericenter of the debris undergoes interruptions of the typical power law $\dot{M} \propto t^{-5/3}$. The time at which the interruptions  occur depends mainly on the distance between the stream of the debris and the secondary.
	
If the TDE occurs in the plane, the initial azimuthal orientation of the orbit can mainly change the time of the first interruption. On the contrary in the case of a TDE perpendicular to the plane, the initial azimuthal orientation of the orbit has no major influences on the first interruption because the average distance between the stream and the secondary is constant. This result confirms the one of \citet{Ricarte2015}.

	\item As theoretically predicted by Equation~\eqref{eq::a_minmax}, for each black hole mass and binary mass ratio, there is an interval out of which the interruptions are not detectable. This interval lies approximately between $0.5$ mpc and $2$ mpc for $M_1 \sim 10^6 \ \msol$.

	\item \label{it::2} The shape of the interruptions are different for different inclinations of the initial orbit of the star. Periodic sharp interruptions of the fallback rate are only present if there are close or direct encounters of the secondary with the stream of the debris. Otherwise, if the stream is mainly perpendicular to the plain, the fallback rate undergoes a first smooth interruption beginning approximately at the theoretical truncation time. Thereafter, a smooth return to the power law can eventually occur but is not necessary present. These results are different than for previous studies using N-body simulations.
    
	\item \label{it::3} In the case $q=0.1$ and $a_{\rm BH}=0.5$ mpc we determined the critical inclinations $\theta_{\rm cr} \sim 70^\circ$ between these two behaviors.
\end{enumerate}

We have not explored here the interesting case where the TDE is around the less massive black hole ($q>1$). In this case the accretion on the secondary, that we did not take into account, could be in the same order than for the primary and could take an important part in the lightcurve.

We have also limited ourselves to parabolic stellar orbits with penetration factors equal to 1. While we do expect the latter condition to be most likely for binary black holes as it is for single black holes, it is not obvious that a parabolic encounter is the most likely outcome of the three-body interaction between the star and the black hole binary. \citet{Coughlin2016} indeed find, based on their three-body calculations, that the energy distribution of the incoming star is relatively broad, ranging from $\approx -2$ to $\approx 2$ in units of $GM/a_{\rm BH}$, where $M$ is the mass of the binary and $a_{\rm BH}$ is the binary separation. While we have not considered the case of non-parabolic encounters in our simulations, we have nonetheless estimated the range of separations for which we expect the binary to alter the fallback rate also in this case (Eq. \ref{eq::a_minmax2}). While for elliptical orbits the range of interesting separations is reduced by a factor of a few at most, very hyperbolic encounters can be substantially affected for in principle much larger separations. 

By looking at observed events, it would be very interesting to check whether some TDE candidates already show the features described in the present paper. The case discussed in \citet{Liu2014} does not have a detailed enough sampling of the lightcurve to conclusively assess its nature as a TDE by binary black holes. A recent very promising case is that of ASASSN-15lh \citep{Leloudas2016}. While this object has been initially considered as a superluminous supernova, \citet{Leloudas2016} make the case that it is instead a TDE. However, the black hole mass inferred based on correlations with the galaxy properties is $\approx 10^8M_{\odot}$, which would then require the black hole to be rapidly spinning in order to disrupt a solar mass star \citep{Kesden2012}. An alternative possibility is that the disrupting black hole is the lower mass companion of a more massive binary system. This interpretation has been proposed by \citet{Coughlin2018}, who also mention in support of this view the fact that the UV lightcurve of ASASSN-15lh does show a sudden dimming and rebrightening after $\approx 100$ days \citep{Leloudas2016}. Note that 100 days is actually the period of a supermassive black hole binary with total mass equal to $10^8M_{\odot}$ and a separation of 1 mpc, at which we expect the fallback to be significantly affected by the binary. \citet{Coughlin2018} show some fallback rate from their large statistical sample of simulated disruptions but do not attempt a direct comparison with the observed lightcurves. One of their fallback curves shows a rebrightening after $\approx$ 100 days but the specific shape does not match closely the observed one. A detailed comparison with our fallback rates is not possible firstly because we do not treat the case of disruptions by the secondary black hole, as mentioned above, and secondarily because we consider lower mass primaries. However, we note that the smooth dimming and rebrightening of ASASSN-15lh is remarkably similar to the fallback curves that we obtain for pole-on disruptions (see for example the middle-right and the lower-middle panels of Fig. 4) as opposed to the sequence of abrupt interruptions for in-plane disruptions. If ASASSN-15lh is indeed a disruption by a binary black hole, we would argue that the stellar orbit must have been significantly inclined with respect to the binary orbit. Another interesting object is iPTF16fnl \citep{Blagorodnova2017}, whose optical lightcurve appears to decline exponentially rather than as $t^{-5/3}$. The interpretation in this case is however much more ambiguous. The signature feature of disruptions by binary black holes is the dimming and then rebrightening of the lightcurve, which is not observed for iPTF16fnl. Actually, \citet{Blagorodnova2017} fit the luminosity evolution with a partial disruption by a single black hole. To date the best case in favour of TDEs by binary black hole remains ASASSN-15lh, although a detailed modeling of this source within this framework has not been carried out yet.


\section*{Acknowledgments}

We wish to thank Guillaume Laibe for valuable remarks on the manuscript. Quentin Vigneron also thanks the \'Ecole Normale Sup\'erieur of Lyon for grant support. We also thank an anonymous referee for a thoughtful report that helped us improving the presentation of our work.


\bibliography{Stage_M1_Notes}

\label{lastpage}
\end{document}